\documentclass[lettersize, 10 pt, conference]{ieeeconf}  
\IEEEoverridecommandlockouts                              
\usepackage{amsmath} % assumes amsmath package installed
\usepackage{amssymb}  % assumes amsmath package installed
\usepackage{bm}
\usepackage{bbm}
\usepackage{graphicx}
\usepackage{color}
\usepackage[font=footnotesize]{subcaption}
\usepackage[font=footnotesize]{caption}
\usepackage{enumerate}
\usepackage{url}
\usepackage{booktabs}
\usepackage{algorithm}
\usepackage{algorithmicx}
\usepackage{algpseudocode}
\usepackage[nobreak]{cite}
\usepackage{comment}
\usepackage{siunitx}
\usepackage{multirow}

\makeatletter
\let\NAT@parse\undefined
\makeatother
\usepackage{hyperref}

\newcommand{\figref}[1]{Fig.~\ref{#1}}
\newcommand{\tabref}[1]{Table \ref{#1}}
\newcommand{\equref}[1]{(\ref{#1})}

\newcommand{\argmax}{\mathop{\rm arg~max}\limits}
\newcommand{\argmin}{\mathop{\rm arg~min}\limits}
\DeclareMathOperator*{\inlineargmax}{\rm arg~max}

\renewcommand{\baselinestretch}{0.973}

\title{\LARGE \bf
Physically Consistent Preferential Bayesian Optimization\\for Food Arrangement
}

\author{Yuhwan Kwon$^{1}$, Yoshihisa Tsurumine$^{1}$, Takeshi Shimmura$^{2, 3}$,\\Sadao Kawamura$^{4}$, and Takamitsu Matsubara$^{1}$% <-this % stops a space
\thanks{$^{1}$YK, YT, and TM are with the 
        Department of Science and Technology, 
        Graduate School of Science and Technology,
        Nara Institute of Science and Technology, Nara, Japan.
        }
\thanks{$^{2}$TS is with the 
        Ganko Food Service Co., Ltd., Osaka, Japan.
        }
\thanks{$^{3}$TS is also with the
        College of Gastronomy Management,
        Ritsumeikan University, Shiga, Japan.
        }
\thanks{$^{4}$SK is with the 
        Department of Robotics,
        Graduate School of Science and Engineering,
        Ritsumeikan University, Shiga, Japan.
        }
\thanks{This work was supported by the Cabinet Office (CAO),
Cross-ministerial Strategic Innovation Promotion Program (SIP),
“An intelligent knowledge processing infrastructure,
integrating physical and virtual domains”
(funding agency: NEDO).}
\thanks{\copyright 2022 IEEE. Personal use of this material is permitted. Permission from IEEE must be obtained for all other uses, in any current or future media, including reprinting/republishing this material for advertising or promotional purposes, creating new collective works, for resale or redistribution to servers or lists, or reuse of any copyrighted component of this work in other works.}
\thanks{
Y. Kwon, Y. Tsurumine, T. Shimmura, S. Kawamura and T. Matsubara, ``Physically Consistent Preferential Bayesian Optimization for Food Arrangement,'' in \textit{IEEE Robotics and Automation Letters (RA-L)}, 2022, doi: 10.1109/LRA.2022.3207560.
}
}
\begin{document}

\maketitle
\thispagestyle{empty}
\pagestyle{empty}

%%%%%%%%%%%%%%%%%%%%%%%%%%%%%%%%%%%%%%%%%%%%%%%%%%%%%%%%%%%%%%%%%%%%%%%%%%%%%%%%
\begin{abstract}
This paper considers the problem of estimating a preferred food arrangement for users from interactive pairwise comparisons using Computer Graphics (CG)-based dish images. As a foodservice industry requirement, we need to utilize domain rules for the geometry of the arrangement (e.g., the food layout of some Japanese dishes is reminiscent of mountains). However, those rules are qualitative and ambiguous; the estimated result might be physically inconsistent (e.g., each food physically interferes, and the arrangement becomes infeasible). To cope with this problem, we propose Physically Consistent Preferential Bayesian Optimization (PCPBO) as a method that obtains physically feasible and preferred arrangements that satisfy domain rules. PCPBO employs a bi-level optimization that combines a physical simulation-based optimization and a Preference-based Bayesian Optimization (PbBO). Our experimental results demonstrated the effectiveness of PCPBO on simulated and actual human users.
\end{abstract}
%%%%%%%%%%%%%%%%%%%%%%%%%%%%%%%%%%%%%%%%%%%%%%%%%%%%%%%%%%%%%%%%%%%%%%%%%%%%%%%%
\section{Introduction}
In recent years, much study has concentrated on using robots in the food industry, especially research on manipulating ingredients and foods\cite{SurveyOnFoodManip, Matsuoka, Takahashi, Endo}. However, most previous studies have focused on robustness\cite{Matsuoka, Takahashi} and accuracy\cite{WangGripper} of the food manipulation; only a few have addressed eye appeal\cite{Endo}. It may be of great value for robots in the foodservice industry to serve dishes with users' preferred food arrangement by providing the estimated preferred arrangement to the grasping and manipulation planners (as in \cite{WangGripper}) that explicitly require target arrangement. Therefore, we focus on the estimation problem of a preferred food arrangement in this paper. 

Comparing pairs of alternatives is intuitively more appealing to humans in determining their preferred one than evaluating them individually, supported by evidence from cognitive psychology\cite{Furnkranz2011}. Preference-based Bayesian Optimization (PbBO)\cite{pbo, pbbo} is one of the methods to estimate preferences efficiently while generating such a two-choice query interactively. Hence, this paper considers the estimation problem of users' preferred food arrangement by PbBO with Computer Graphics (CG)-based dish images. The use of CG is naturally motivated because it is difficult to generate those many images for comparison in a real environment\cite{Matsuoka}.

%%%%%%%%%%%%========================================================%%%%%%%%%%%%
\begin{figure}[t]
 \begin{center}
  \includegraphics[width=0.95\linewidth, trim=0 0 30 0, clip]{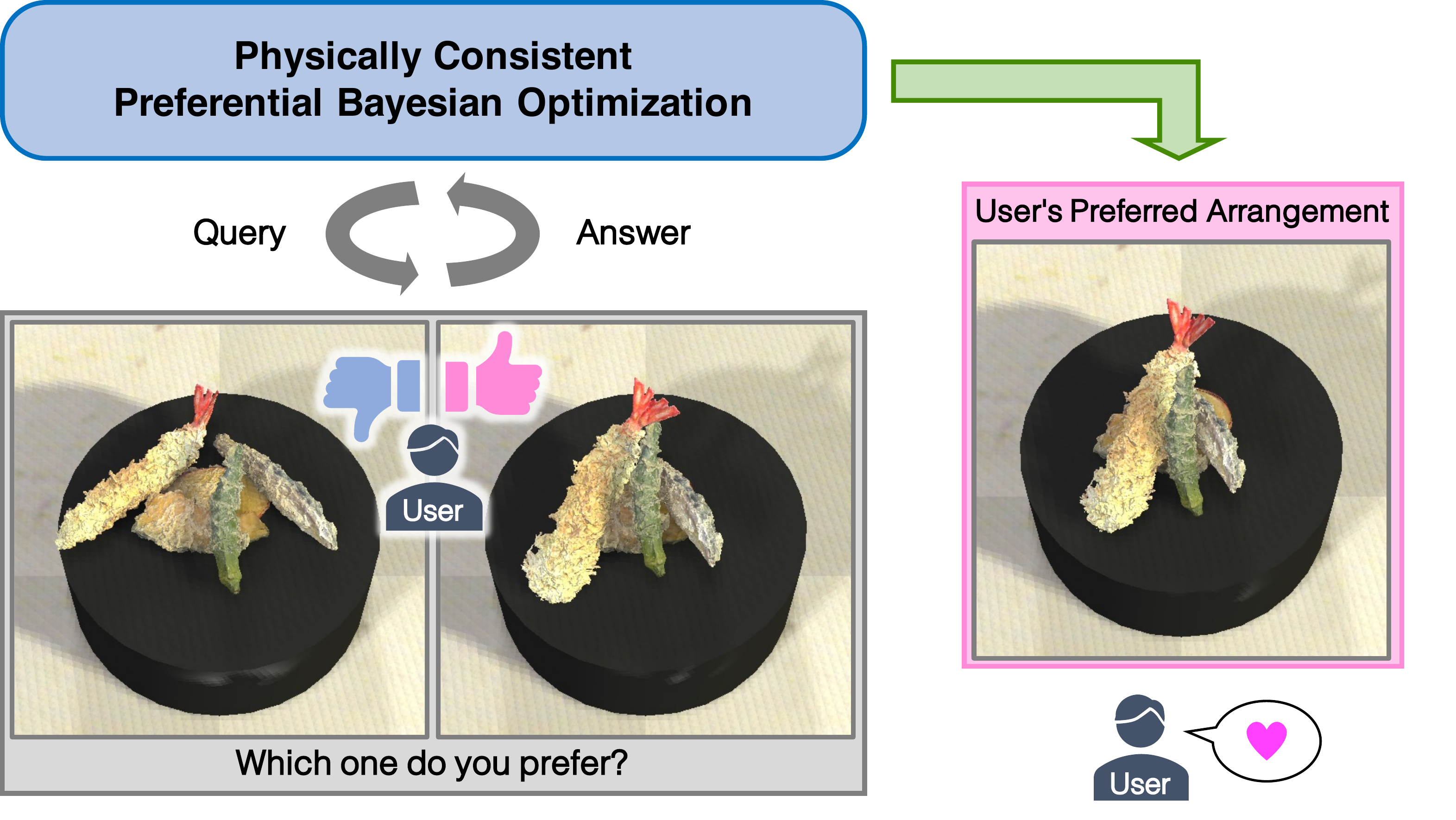}
  \caption{Rough sketch of preferred food arrangement estimation by proposed Physically Consistent Preferential Bayesian Optimization, which estimates a preferred arrangement that satisfies both physical laws and domain rules through interactive pairwise comparisons.}
  \label{fig: top_img}
 \end{center}
\end{figure}
%%%%%%%%%%%%========================================================%%%%%%%%%%%%

As a foodservice industry requirement, we need to utilize {\it domain rules} about the geometry of the arrangement as used by chefs\cite{Miyazawa}. However, those rules are qualitative and ambiguous; the arrangements based on the rules can be {\it physically inconsistent} (e.g., physical interference of foods on CG makes the arrangement physically infeasible). For example, in Japanese dishes as tempura, the rule ``place the foods to resemble mountains'' is essential\cite{Miyazawa}. However, no specific instructions are given on how to place them without collapsing. Therefore, for robotic automation purposes, it becomes crucial to estimate the preferred arrangement that is consistent with both such rules and physical constraints.
 
Since the domain rules are geometric, they can be easily considered in the food image generation so that PbBO can estimate the food arrangement that satisfies the domain rules and user preferences. However, physical consistency cannot be considered simultaneously in the same manner due to its complexity. To overcome this issue, we utilize {\it physics simulation-based optimization}; by generating dish images that satisfy both domain rules and physical consistency through the optimization, we could estimate user-preferred food arrangements that meet both of them.

In this paper, we propose Physically Consistent Preferential Bayesian Optimization (PCPBO) as a method that estimates the preferred food arrangement satisfying both physical laws and domain rules. For its estimation, PCPBO employs a bi-level optimization that combines a physical simulation-based optimization and PbBO to generate two-choice arrangements consistent with them. Figure~\ref{fig: top_img} shows how PCPBO estimates it through interactive pairwise comparisons with a user. We also developed an environment of physical simulation with CG on Japanese food arrangement tasks and experimentally evaluated our proposed method. Our experimental results on the tasks demonstrated the effectiveness of PCPBO for estimating physically consistent and users' preferred arrangements to simulated and actual human users.

The following are this paper's main contributions:
\begin{itemize}
    \item To our knowledge, we proposed the first framework for estimating the preferred food arrangement that considers the geometric rules and physical constraints essential for robotic food arrangement in the foodservice industry. 
    \item We formulated PCPBO as the solution method to the above problem with bi-level optimization.
    \item We conducted experiments on tasks that reflect the rules of Japanese cuisine for real and simulated users.
\end{itemize}

%%%%%%%%%%%%%%%%%%%%%%%%%%%%%%%%%%%%%%%%%%%%%%%%%%%%%%%%%%%%%%%%%%%%%%%%%%%%%%%%
\section{Related Work}
%%%%%%%%%%%%%%%%%%%%%%%%%%%%%%%%%%%%%%%%%%%%%%%%%%%%%%%%%%%%%%%%%%%%%%%%%%%%%%%%
\subsection{Studies for Manipulating Foods}
Many studies have addressed food recognition\cite{SurveyOnFoodComp} and handling mechanisms\cite{SurveyOnFoodManip}, although few previous studies focused on food-manipulation planning. Matsuoka \textit{et al.}\ focused on the arrangement errors and environmental changes in the arrangement of tempura, a popular Japanese dish, and proposed a food arrangement planning method that is robust to such errors\cite{Matsuoka}. Takahashi \textit{et al.}\ proposed a method that utilizes the uncertainty of deep learning models obtained through self-supervised learning to grasp granular ingredients from a tray as accurately as possible\cite{Takahashi}. While those studies focused on robustness and accuracy, other studies focused on preference. Endo \textit{et al.}\ developed a gripper that arranges noodles and chopped ingredients in a conical shape to make them more appealing and verified their performance under several conditions\cite{Endo}. Junge \textit{et al.}\ proposed a method using Bayesian optimization (BO) to optimize the taste and the preference of omelets cooked by a robot\cite{robocook_bbo}. Our study differs because it considers how foods are arranged to match the preference of the appearance.

%%%%%%%%%%%%%%%%%%%%%%%%%%%%%%%%%%%%%%%%%%%%%%%%%%%%%%%%%%%%%%%%%%%%%%%%%%%%%%%%
\subsection{Preference Learning}
Several studies have attempted to optimize CG and animation parameters based on human preference. Eric \textit{et al.}\ proposed an Active Preference Learning method with BO to estimate the desired parameters for changing the appearance of materials\cite{Brochu}. Koyama \textit{et al.}\ proposed an original User Interface and an optimization method to adjust image and CG parameters\cite{Koyama}. Several studies also leverage Preference Learning in the context of robotics. Sadigh \textit{et al.}\ proposed an active learning method to learn human reward functions from pairwise comparisons using Gaussian Process (GP)\cite{Sadigh2017ActivePL}. Christiano \textit{et al.}\ also proposed a method of learning reward functions in Deep Reinforcement Learning from pairwise comparisons\cite{Christiano}. Roveda \textit{et al.}\ proposed a method to tune the parameters of a path-based velocity planner for a sealing task with preference-based optimization\cite{pbo_dmp}. There are also several PbBO studies on a framework for machine learning (not for robotics). Gonzalez \textit{et al.}\ proposed a BO method for pairwise comparison and an improved acquisition function by focusing on the characteristics of the bandit problem\cite{pbo}. Siivola \textit{et al.}\ proposed a PbBO method under batched query settings and discussed some approximate inferences\cite{pbbo}. Our study differs because it can be applied to tasks where ambiguous rules and physical consistency need to be considered, such as food arrangement. In addition, considering the users' burden and maintaining motivation for answering, we adopted a sample-efficient Bayesian optimization framework. In the experiments shown later, the number of responses required in our approach was several dozed. On the other hand, deep learning-based methods may require at least several hundred responses to find a preferred arrangement.
%%%%%%%%%%%%%%%%%%%%%%%%%%%%%%%%%%%%%%%%%%%%%%%%%%%%%%%%%%%%%%%%%%%%%%%%%%%%%%%%
\section{Preference-based Bayesian Optimization}\label{subsec: pbbo}
By following \cite{pbbo}, a Bayesian optimization (BO)\cite{PracticalBO} method finds ${\bm w}^* = \inlineargmax_{{\bm w}} f({\bm w})$ from as few queries as possible for black-box function $f({\bm w}): {\mathcal W} \rightarrow {\mathbb R}$, which is difficult to evaluate directly. Therefore, BO first constructs a surrogate model that is relatively easy to evaluate with a probabilistic model and then regresses $f$ on it from set of queries $\{{\bm w}_i \}_{i=0}^{K-1}$ and corresponding $\{f({\bm w}_i) \}_{i=0}^{K-1}$. Then BO selects new query ${\bm w}'$ by evaluating the surrogate with an acquisition function that considers the balance between exploration and exploitation.

On the other hand, in PbBO, although we cannot directly observe $f({\bm w})$, instead, answer $y$ from a user corresponding to two-choice query $({\bm w}^0, {\bm w}^1)$ is given by the magnitude relation between them: 
\begin{align}
    y = \left\{
        \begin{array}{ll}
        0, & \text{if $f({\bm w}^0) \geq f({\bm w}^1)$}\\
        1, & \text{if $f({\bm w}^0) < f({\bm w}^1)$}
        \end{array}
        \right..
        \label{equ: ideal_response}
\end{align}
Next we determine a candidate point for a new query using $p({\bm f} | {Y})$, which is the probability distribution of ${\bm f}:=f(W)$ when user's answers $Y:=\{y_i\}_{i=0}^{K-1}$ are given to queries $W:= \{ {\bm w}^0_i, {\bm w}_i^1 \}_{i=0}^{K-1}$. Here we simplify the notation by writing $f({\bm w}^0)$ and $f({\bm w}^1)$ as $f^0$ and $f^1$ and the preference relations as ${\bm w}^0 \succ {\bm w}^1 := {f^0} \geq {f^1}$ and ${\bm w}^0 \prec {\bm w}^1 := {f^0} < {f^1}$.

Let the hyperparameter be ${\bm \theta} \in {\Theta}$, and let prior distribution $p({\bm f}|{\bm \theta})$ be the Gaussian Process (GP)\cite{gpml} with mean $\bm 0$ and covariance matrix $\mathbf K$. Furthermore, assuming that the preference relation for a given two-choice query is polluted by Gaussian noise, the likelihood of the magnitude relation \cite{Chu2005} can be defined as $p({\bm w}^0 \prec {\bm w}^1 | f^0, f^1) = \int_{-\infty}^{\frac{f^1-f^0}{\sqrt{2}\sigma}}\mathcal{N}(\gamma|0, 1)d\gamma$. Also, for $Y$, it is defined as $p({Y}|{\bm f}) = \prod_{i=0}^{K-1} p({\bm w}^0_i \prec {\bm w}^1_i | f^0_i, f^1_i)$. From the Bayes' theorem, posterior distribution is defined as $p({\bm f}|{Y}, {\bm \theta}) \propto p({Y}|{\bm f})p({\bm f}|{\bm \theta})$. Since we can only obtain the likelihood numerically, we cannot obtain the posterior analytically. Therefore, we approximate $p({\bm f}|{Y}, {\bm \theta})$ by the Variational Bayesian method \cite{vi_revisit}. If the variational distribution is $q({\bm f}) \approx p({\bm f}|{Y}, {\bm \theta})$, the inference of $f'$ corresponding to new candidate point ${\bm w}'$ is given by $p(f'| {Y}, {\bm w}') = \int p(f'|{Y}, {\bm f}, {\bm w}')q({\bm f}) d{\bm f}$. Using this, we can construct the acquisition function for selecting a candidate point for a new query.

%%%%%%%%%%%%%%%%%%%%%%%%%%%%%%%%%%%%%%%%%%%%%%%%%%%%%%%%%%%%%%%%%%%%%%%%%%%%%%%%
\section{Approach}
%%%%%%%%%%%%%%%%%%%%%%%%%%%%%%%%%%%%%%%%%%%%%%%%%%%%%%%%%%%%%%%%%%%%%%%%%%%%%%%%
\subsection{Problem Formulation}\label{subsec: prob_form}
Let $M$ be the number of foods to be arranged, let $n_s$ be the number of dimensions of each food's state, let ${\bm s}_m \in {\mathcal S} \subseteq {\mathbb R}^{n_s \times M}$ be the continuous state of all the foods when the $m < M$th food is set, and let $\tau := ({\bm s}_0, \dots, {\bm s}_{M})$ be the state sequence. Next we assume that we can evaluate the human preference of the food arrangement from image ${\bm x}_M \in {\mathcal X} \subset {\mathbb R}^{n_{\bm x}}$ corresponding to final dish state ${\bm s}_M$, where $n_x := \dim {\bm x}$. We simplify the notation by referring to ${\bm s}_M$ and ${\bm x}_M$ as ${\bf s}$ and ${\bf x}$. Hence, using maps $k: {\mathcal S} \rightarrow {\mathcal X}$ and $g: {\mathcal X} \rightarrow {\mathbb R}$ and composite map $h:=g \circ k$, the degree of human preference for ${\bf s}$ is given by $c = h({\bf s})$.
Here $g$ is interpreted as user's latent preference function, but we define $h$ as the preference function for convenience.

Let $\mathcal{R} \subset \mathcal {S}$ be set of final dish states that satisfy the domain rules used by chefs; we estimate user's preferred arrangement ${\bf s}^* = \inlineargmax_{{\bf s} \in \mathcal{R}} h({\bf s})$. However, since ${\bf s}$ can be high-dimensional and it is challenging to perform optimization directly in that space, we introduce a low-dimensional parameter ${\bm w} \in {\mathcal W}\subseteq {\mathbb R}^{n_{\bm w}}$ (e.g., a part of a particular food item), where $n_{\bm w} := \dim {\bm w}$. In addition, we design an injective function $d: \mathcal{W} \rightarrow \mathcal{R}$ to obtain a variety of rule-abiding arrangements that change with $\bm w$. Finally, we replace the problem of estimating ${\bf s}^*$ with that of estimating preferred weight ${\bm w}^*$:
\begin{equation}
    {\bm w}^* = \argmax_{{\bm w}} h(d({\bm w})).
    \label{equ: persnal_w}
\end{equation}

%%%%%%%%%%%%========================================================%%%%%%%%%%%%
\begin{figure}[t]
 \begin{center}
  \includegraphics[width=0.85\linewidth, trim=0 0 0 0, clip]{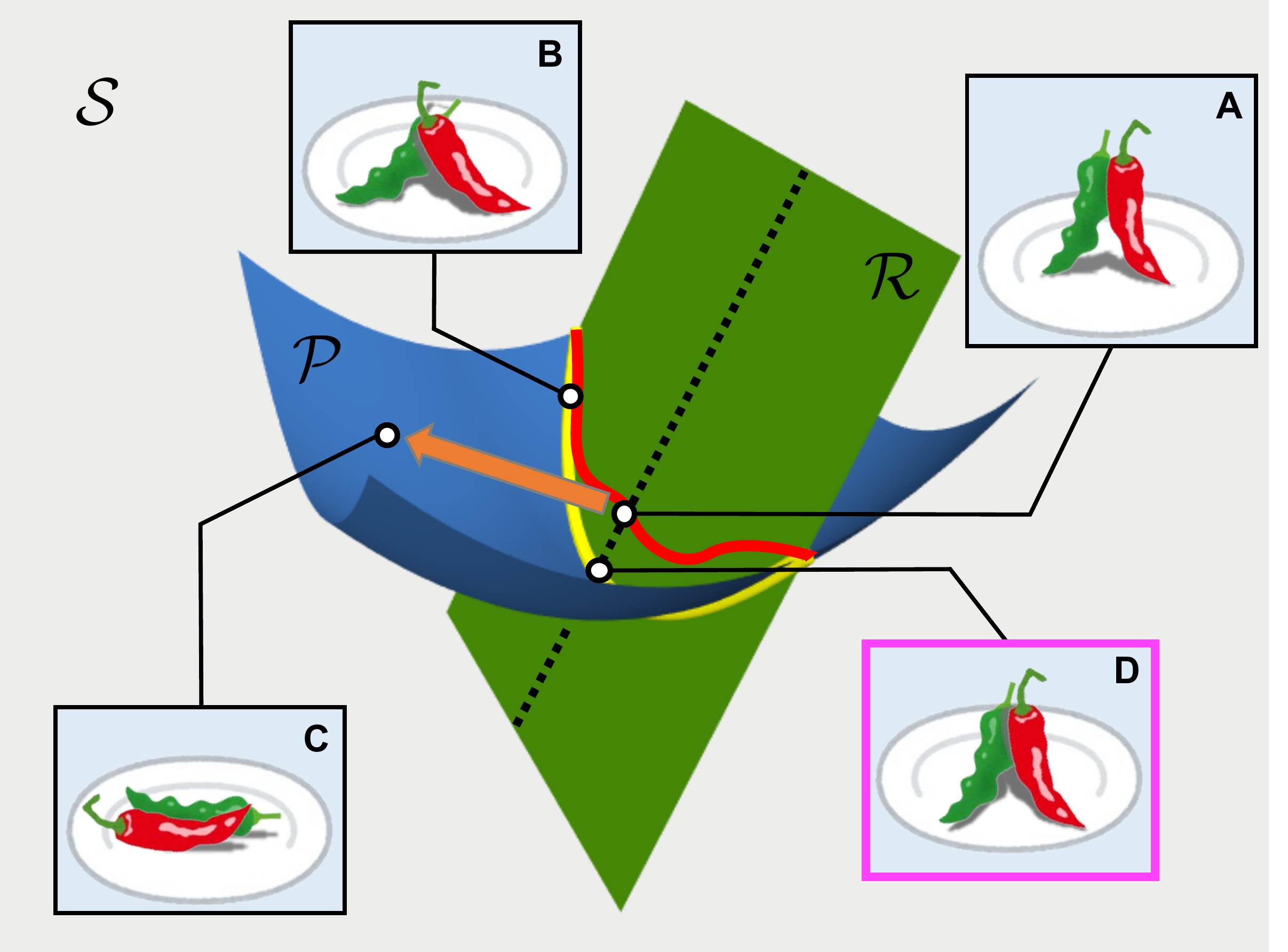}
  \caption{Conceptual diagram of a space satisfying physical consistency (blue) and a space following domain rules (green) in food arrangement state space: Our proposed PCPBO generates arrangements in the yellow curve where both the rules and physical consistency are satisfied. Also, the image highlighted in pink is what we aim to estimate.}
  \label{fig: our_insight}
 \end{center}
\end{figure}
%%%%%%%%%%%%========================================================%%%%%%%%%%%%

%%%%%%%%%%%%%%%%%%%%%%%%%%%%%%%%%%%%%%%%%%%%%%%%%%%%%%%%%%%%%%%%%%%%%%%%%%%%%%%%
\subsection{Idea for Solution}
We use a CG-based approach to generate $\bf x$ corresponding to $\bm w$ and aim to solve \equref{equ: persnal_w} with PbBO. However, ${\bf s}^*$ estimated from comparisons that consider only the rules can be physically infeasible. Now we explain this problem using \figref{fig: our_insight}, which shows state space $\mathcal S$ regarding the arrangement of red and green peppers, and the green and blue regions represent $\mathcal R$ and $\mathcal P$, the set of physically consistent arrangements. In this example, we set the following rule: ``Place the foods to resemble mountains.'' Furthermore, the red curve represents $d$, and $\bm w$ corresponds to the angle between the two peppers. 

In \figref{fig: our_insight}, \textsf{A} and \textsf{B} have different values of $\bm w$, so the angles are different from each other. If we consider \textsf{A} corresponds to ${\bm w}^*$, we can represent it as $d({\bm w}^*)$ because it lies on the red curve. Here the black dashed line represents the set of arrangements corresponding to ${\bm w}^*$. Since we are interested in physically arranging the foods, we place them with \textsf{A} as a reference. Unfortunately, because of $\mathcal{R} \ni \textsf{B} \notin \mathcal{P}$, we get \textsf{C} where collapses and does not follow the rule, shown at the tip of the orange arrow. This means that physically executing such an estimated arrangement would result in an unexpected one, which might not satisfy the rule. Thus, just using the rules fails because they are qualitative and ambiguous, and this tendency is especially pronounced in food arrangements where multiple objects come into contact.

Our idea to tackle this difficulty is to generate the arrangements simultaneously considering the physical consistency and the domain rules in PbBO for solving \equref{equ: persnal_w}. In \figref{fig: our_insight}, the yellow curve represents the space where both the consistency and rules are satisfied. Hence, \textsf{B} and \textsf{D} are physically feasible, and \textsf{D} at the intersection with the dashed line is the arrangement we aim to estimate. Note that we describe the arrangement on the yellow curve as $\hat{\bf s}$, and when we want to emphasize that it corresponds to $\bm w$, we write it as $\hat{\bf s}_{\bm w}$.

%%%%%%%%%%%%%%%%%%%%%%%%%%%%%%%%%%%%%%%%%%%%%%%%%%%%%%%%%%%%%%%%%%%%%%%%%%%%%%%%
\subsection{Physically Consistent Preferential Bayesian Optimization}

We propose Physically Consistent Preferential Bayesian Optimization (PCPBO) to implement our idea. PCPBO performs alternately with a model-based optimization to optimize the arrangements against cost function $\| {\bf s} - d({\bm w}) \|$ and PbBO to estimate ${\bm w}^*$ in the preference function based on the optimized arrangement. Let ${\bm a}_m \in {\mathcal A} \subseteq {\mathbb R}^{n_a}$ be the continuous action of arranging an $m$th food on a dish, and let $U := ({\bm a}_0, \dots, {\bm a}_{M-1})$ be the action sequence. Then ${\bf s}$ is obtained by recursively applying $U$ to the state transition model with physical consistency $F: {\mathcal P} \times {\mathcal A} \rightarrow {\mathcal P}$. Therefore, we use ${\bf s}$ and $\| {\bf s} - d({\bm w}) \|$ in the model-based optimization to obtain optimal action sequence $\hat{U}$. We use notations $\hat{\bf s}_{\bm w}$ and $\hat{U}_{\bm w}$ to emphasize their dependence on ${\bm w}$.

To summarize, our PCPBO is formulated as the following bi-level optimization:
\begin{align}
    {\bm w}^* &= \argmax_{\bm w} h(\xi(\hat{U}_{\bm w})) \hspace{3mm} \mathrm{with} \label{equ: upper}\\
    \hat{U}_{\bm w} &= \argmin_{U} \| {\bm s}_M - d({\bm w}) \|, \label{equ: lower}\\
    {\rm s.t.}\; {\bm s}_{m+1} &= F({\bm s}_m, {\bm a}_m), {\bm s}_0 = {\bm \alpha}, \nonumber
\end{align}
where $\xi$ is a map that applies $F$ recursively to give corresponding $\hat{\bf s}_{\bm w}$ from $\hat{U}_{\bm w}$ and ${\bm \alpha}$ is the initial state. 

\begin{figure}[t]
 \begin{center}
  \includegraphics[width=\linewidth, trim=0 0 0 0, clip]{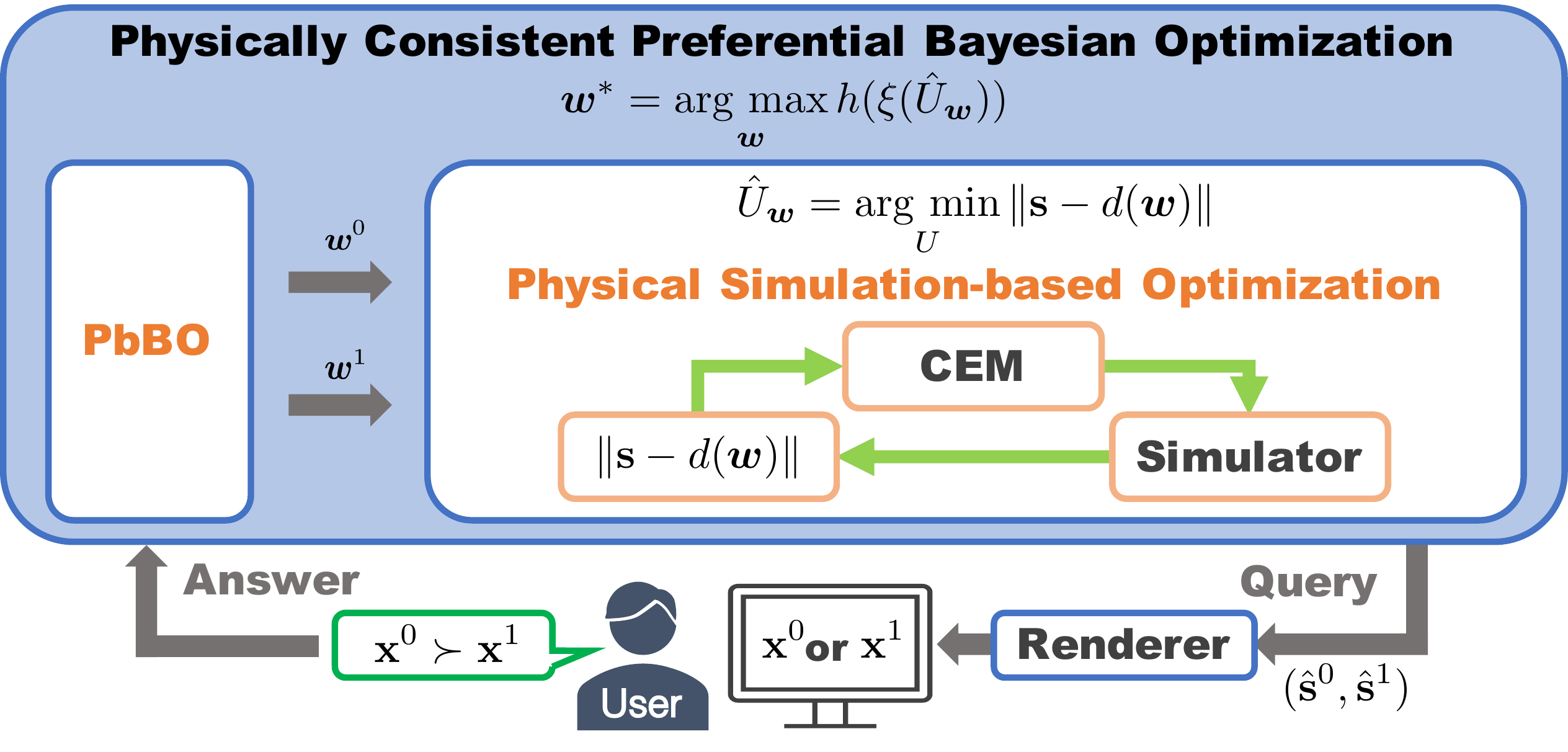}
  \caption{Framework of proposed Physically Consistent Preferential Bayesian Optimization (PCPBO)}
  \label{fig: pcpbo}
 \end{center}
\end{figure}

Since this bi-level optimization problem cannot be solved analytically, we solve it by the following sequential procedure with a query generation that simultaneously considers the physical laws and domain rules:
\begin{enumerate}
    \item select $({\bm w}^0, {\bm w}^1)$ by an acquisition function (\ref{subsec: acq_fcn});
    \item generate $(\hat{\bf s}^{0}, \hat{\bf s}^{1})$ corresponding to $({\bm w}^0, {\bm w}^1)$ by physical simulation-based optimization (\ref{subsec: opt_xi});
    \item generate $({\bf x}^{0}, {\bf x}^{1})$ corresponding to $(\hat{\bf s}^{0}, \hat{\bf s}^{1})$ by a renderer;
    \item present $({\bf x}^{0}, {\bf x}^{1})$ as a query to a user;
    \item do a query synthesis if the user skipped (\ref{subsec: query_synth});
    \item do learning preference function using answer $y$ and $({\bm w}^0, {\bm w}^1)$ (\ref{subsec: pbbo}),
\end{enumerate}
where $\hat{\bf s}_{{\bm w}^k}$ and ${\bf x}_{{\bm w}^k}$ are denoted as $\hat{\bf s}^{k}$ and ${\bf x}^{k}$ for simplicity of notation. Here steps 1) through 5) are repeated until the estimation converges. We illustrate the flow of this procedure in \figref{fig: pcpbo}. Also, the code is available on our project page\footnote{\url{https://y-kwon.github.io/pcpbo/}}. From the following, we describe each procedure. 

%%%%%%%%%%%%%%%%%%%%%%%%%%%%%%%%%%%%%%%%%%%%%%%%%%%%%%%%%%%%%%%%%%%%%%%%%%%%%%%%
\subsection{Weight Candidate Selection}\label{subsec: acq_fcn}
As in the previous study\cite{pbo}, the space of ${\bm w}$ is discretized, and Thompson sampling is used to determine candidate points $({\bm w}_0, {\bm w}_1)$ for a new query. Specifically, after $f^0, f^1 \sim q({\bm f})$, ${\bm w}_0$ and ${\bm w}_1$ are searched for ${\bm w}^0 = \inlineargmax_{\bm w} f^0$ and ${\bm w}^1 = \inlineargmax_{\bm w} f^1$, respectively. However, we found a tendency to present candidates with close distances on $\bm w$ when $\dim {\bm w}$ is low. Therefore, we finitely reselect the candidates when their distances are close. In addition, they are randomly selected for $N_{\mathrm{init}}$ times in the early stages. Furthermore, we design the $\bm w$ space so that each dimension is within $[0, 1]$.

%%%%%%%%%%%%%%%%%%%%%%%%%%%%%%%%%%%%%%%%%%%%%%%%%%%%%%%%%%%%%%%%%%%%%%%%%%%%%%%%
\subsection{Physical Simulation-based Optimization}\label{subsec: opt_xi}
Designing $d$ so that $d({\bm w})$ satisfies the physical constraints over $\mathcal W$ is difficult since $d$ is designed based on chefs' qualitative and ambiguous advice, and there are complex contacts between objects in the food arrangements. Hence, we seek a physically consistent ${\bm s}_M$ that is sufficiently similar to $d({\bm w})$ that meets the rules but does not consider physical feasibility. PCPBO finds the optimal action sequence $\hat{U}$ that generates the desired arrangement by solving an optimal control problem with $\| {\bf s} - d({\bm w}) \|$ as the cost function in \equref{equ: lower} using physical simulation-based optimization. Since the state transitions become nonlinear and complex when multiple foods come into contact, \equref{equ: lower} is optimized by the Cross Entropy Method (CEM)\cite{mppi}, a sampling-based optimal control method. Also, we constructed a parallelized food arrangement simulation environment on Isaac Gym\cite{isaac} and directly used it as the state transition model for CEM.

Since a physical simulator is used directly as a state transition model, PCPBO does not require learning state transitions in advance to handle food items with different colors and shapes but only preparing their meshes. Specifically, we prepared meshes for rendering, used for user queries, and meshes for detecting contact. The former was obtained by 3D scanning food samples, and the latter was designed with a simplified shape (e.g., rectangular) based on the former. In addition, the design of $d$ and $\bm w$ for new dishes is performed on CG without a physics engine using a mesh for rendering\cite{vrep}. Note that for tasks that do not require consideration of physical constraints, much performance difference would not occur between PCPBO and PbBO since there is no need to perform physical simulation-based optimization.

%%%%%%%%%%%%========================================================%%%%%%%%%%%%
%%%%%%%%%%%%========================================================%%%%%%%%%%%%

%%%%%%%%%%%%%%%%%%%%%%%%%%%%%%%%%%%%%%%%%%%%%%%%%%%%%%%%%%%%%%%%%%%%%%%%%%%%%%%%
\subsection{Query Skipping and Synthesis}\label{subsec: query_synth}
We believe that if both $\hat{\bf s}^{0}$ and $\hat{\bf s}^{1}$ are highly undesirable, the users find it difficult to answer and might make a wrong selection. Therefore, we asked them to skip the query in this case. On the other hand, to effectively use the skipped queries, we synthesized new queries using them and the arrangement selected in the past and added them to the dataset. More precisely, the two arrangements in the skipped query are combined with the most recently selected arrangement to make two queries. We call this method query synthesis. If the selected arrangement does not yet exist, we accumulate the queries until they are no longer skipped.

However, since ${\bf s}$ can be high-dimensional, we design map $d: \mathcal{W} \rightarrow \mathcal{R}$ by introducing lower-dimensional weight ${\bm w} \in {\mathcal W}\subseteq {\mathbb R}^{n_{\bm w}}$ that satisfies the rules, although it can also change the arrangement, where $n_{\bm w} := \dim {\bm w}$.

%%%%%%%%%%%%%%%%%%%%%%%%%%%%%%%%%%%%%%%%%%%%%%%%%%%%%%%%%%%%%%%%%%%%%%%%%%%%%%%%
\section{Simulation Experiment}
%%%%%%%%%%%%%%%%%%%%%%%%%%%%%%%%%%%%%%%%%%%%%%%%%%%%%%%%%%%%%%%%%%%%%%%%%%%%%%%%
\subsection{Experimental Setup}
We conducted a simulation experiment with simulated users that automatically answers queries according to designed human preference models. Note that we assume that no access to their preference models is given; instead, we estimate $\bm w$ in the models from the answers according to those models. This experiment has two purposes. First, we investigated the estimation performance of the preferred weights with and without considering physical consistency using simulated users that answers accurately. Second, we investigated the effectiveness of the query synthesis using the users that returned uncertain answers.

%%%%%%%%%%%%%%%%%%%%%%%%%%%%%%%%%%%%%%%%%%%%%%%%%%%%%%%%%%%%%%%%%%%%%%%%%%%%%%%%
\subsubsection{Arrangement Tasks}
\begin{figure}[t]
  \centering
  \begin{minipage}{0.485\linewidth} % {0.49\linewidth}
    \centering
    \includegraphics[clip, trim=0 0 0 0, width=0.9\linewidth]{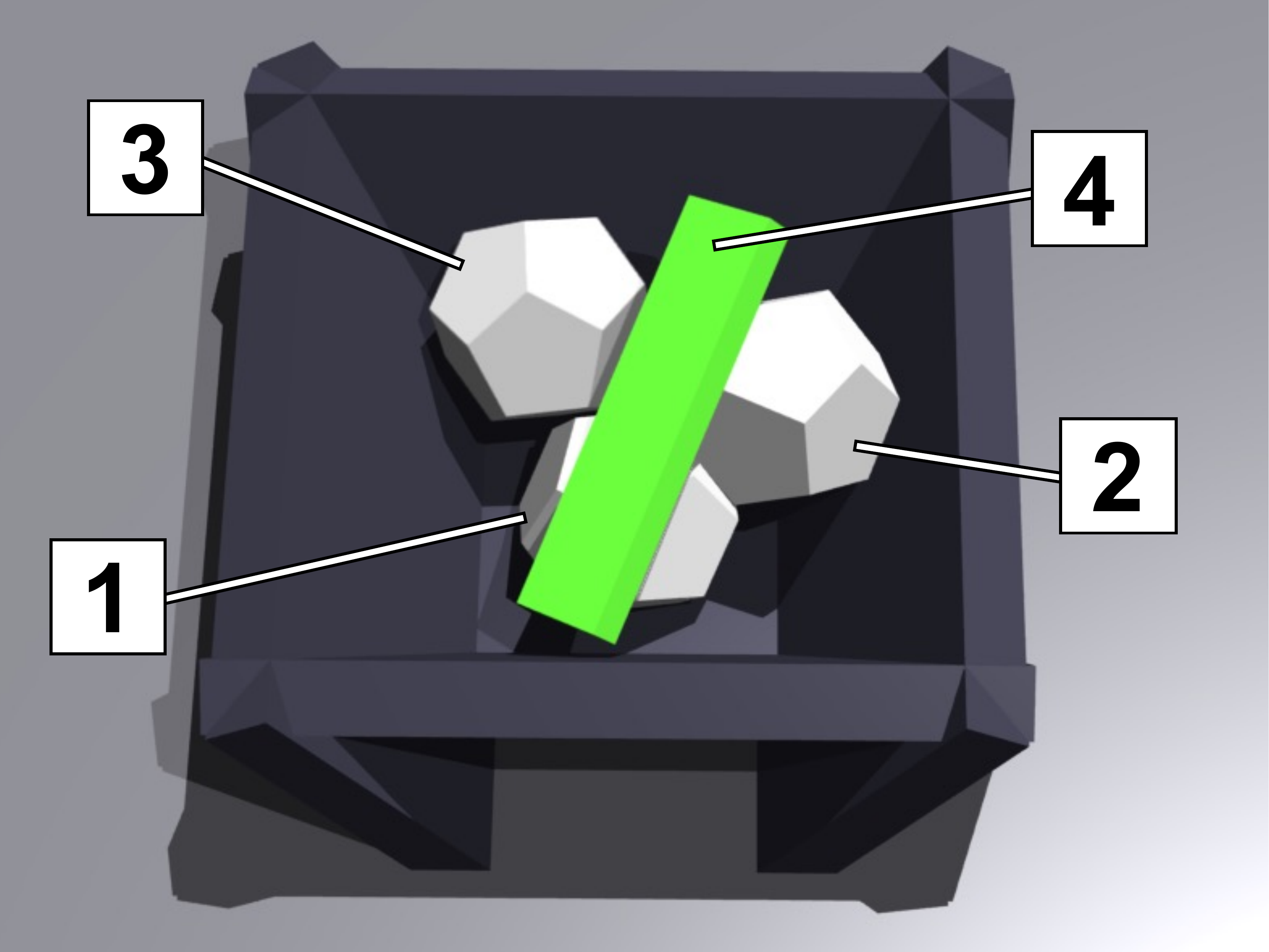}
    \subcaption{Simmered taro}
    \label{fig: taro}
  \end{minipage}
  \begin{minipage}{0.485\linewidth} % {0.49\linewidth}
    \centering
    \includegraphics[clip, trim=0 0 0 0, width=0.9\linewidth]{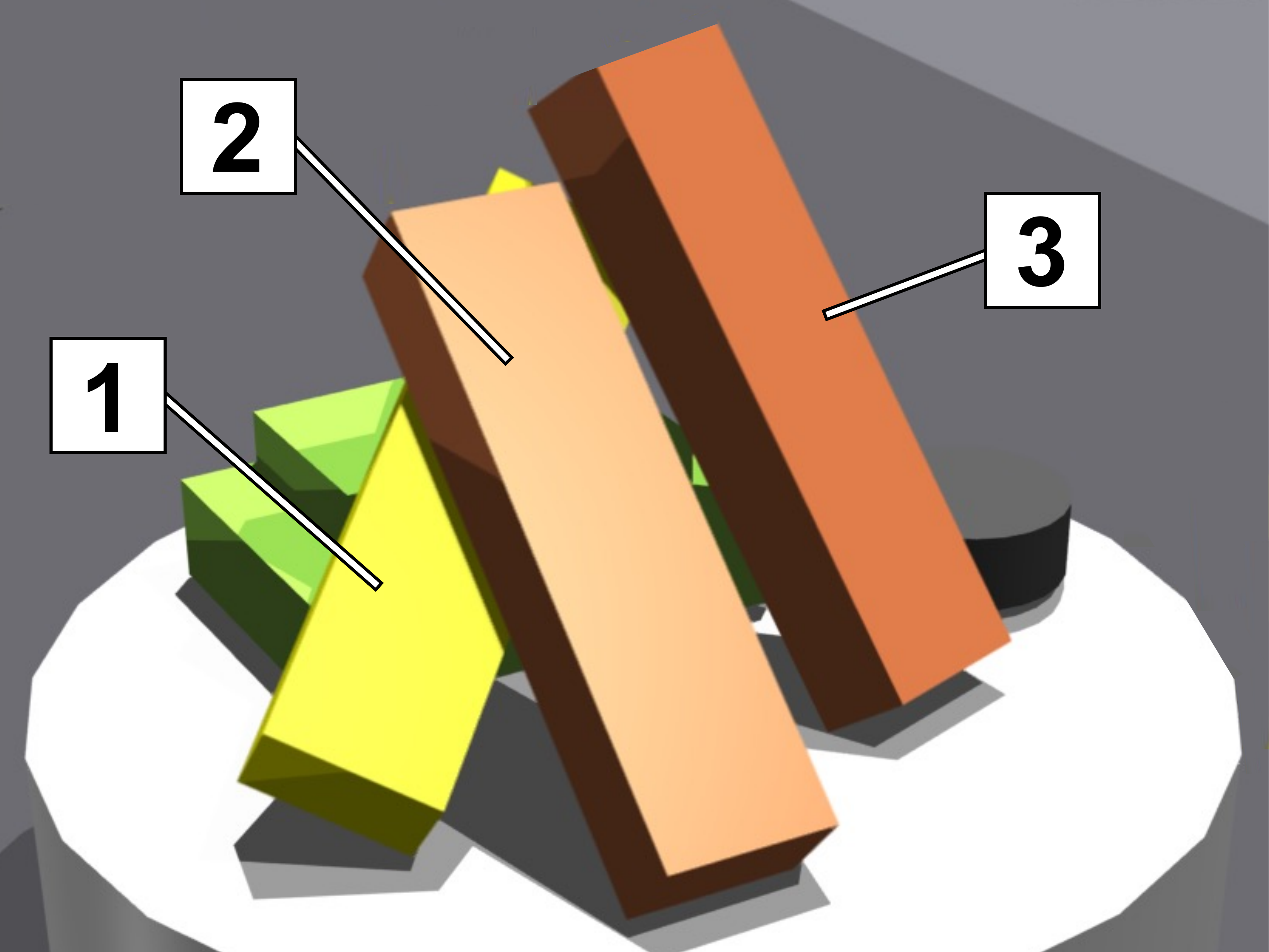}
    \subcaption{Deep-fried shrimp}
    \label{fig: ebifry}
  \end{minipage}
  \caption{Physical contact environment in the simulation experiments: We approximate foods and the dish with rigid bodies of simple shapes. The numbers in the figure indicate the order of arrangement. Also, each rigid body is colored for debugging purposes only.}
  \label{fig: sim_env}
\end{figure}

We chose the arrangement of simmered taro and deep-fried shrimp as the tasks for our simulation experiment. Those were designed from rules based on advice from the professional chefs of a Japanese foodservice company. These rules are qualitative and ambiguous (\tabref{tab: knowledge}). Figure~\ref{fig: sim_env} shows the physical contact environment with Isaac Gym\cite{isaac}. In \figref{fig: taro}, the white and green object represents taro and snap pea. In \figref{fig: ebifry}, the yellow and orange cuboids represent the fried shrimp, and the green and gray rigid bodies represent shredded cabbage and a small dipping bowl, each of which is fixed on a white plate. Note that we omit the environment for visualization and use only that for contact detection since simulated users (described below) can directly observe the food item poses.

In both tasks, we defined ${\bm s}_m$ and ${\bm a}_m$ as
\begin{align}
    {\bm s}_m &= (x_0, y_0, z_0, \theta_0, \phi_0, \psi_0, \ldots, \psi_M), \label{equ: exp_state}\\
    {\bm a}_m &= (x_m^{\mathrm{ref}}, y_m^{\mathrm{ref}}, \psi_m^{\mathrm{ref}}), \label{equ: exp_action}
\end{align}
where $(x, y, z)$ and $(\theta, \phi, \psi)$ are the position and posture (roll, pitch, yaw angles) of the food, and $\mathrm{ref}$ stands for reference. Note that we set $\theta_m^{\mathrm{ref}}$ and $\phi_m^{\mathrm{ref}}$ to 0, and $z_m^{\mathrm{ref}}$ is determined from a heuristic algorithm that drops the food from as low a height as possible. In simmered taro environment, for the four food items to appear, state sequence $\tau = ({\bm s}_0, {\bm s}_1, {\bm s}_2, {\bm s}_3)$. Since we also give a specific action to the taros, action sequence $U=({\bm a}_3)$, which is restricted by the rules. In deep-fried shrimp environment, we fix all the foods except the fried shrimp, so $\tau = ({\bm s}_0, {\bm s}_1, {\bm s}_2)$, but since we give a specific action to the fried shrimp, represented by the yellow cuboid, $U=({\bm a}_1, {\bm a}_2)$, which is also restricted by the rules.

%%%%%%%%%%%%========================================================%%%%%%%%%%%%
%%%%%%%%%%%%========================================================%%%%%%%%%%%%
%%%%%%%%%%%%========================================================%%%%%%%%%%%%
\begin{table}[t]
\centering
  \caption{Examples of qualitative and ambiguous rules in Japanese cuisine\cite{Miyazawa}: The chefs utilize several of these rules depending on the dish.}
  \begin{tabular}{c||l} \hline
    \textbf{No.} & Description \\ \hline \hline
    \textbf{1} & View a dish from an angle. \\ \hline
    \textbf{2} & The colors of the dish and the foods are complementary. \\ \hline
    \textbf{3} & Place the foods to resemble mountains. \\ \hline
    \textbf{4} & Place the foods low in the front and high in the back. \\ \hline
    \textbf{5} & Pay attention to the placement of vivid color foods. \\ \hline
  \end{tabular}
  \label{tab: knowledge}
\end{table}

%%%%%%%%%%%%========================================================%%%%%%%%%%%%

%%%%%%%%%%%%%%%%%%%%%%%%%%%%%%%%%%%%%%%%%%%%%%%%%%%%%%%%%%%%%%%%%%%%%%%%%%%%%%%%
\subsubsection{Preferred Weight Estimation Methods}
To achieve the first purpose, we prepared PbBO as a comparison method, which does not consider physical consistency but only the rules. This PbBO physically places the foods with $d({\bm w})$ as a reference arrangement. In simmered taro environment, we assume that one-dimensional $\bm w$ space corresponds to $\psi_3$ and that $x$ and $y$ are roughly determined from $\psi$ by the rules. In deep-fried shrimp environment, both methods search for two-dimensional $\bm w$ space corresponding to $\psi_1$ and $\psi_2$, assuming that the rules roughly determine $x$ and $y$ from $\psi$ as well. Each dimension of the $\bm w$ space is discretized at 21 points. We also performed an ablation study in the deep-fried shrimp environment to investigate the effectiveness of query synthesis, which is the second purpose of this experiment. There are three cases: query synthesis is enabled, it is disabled, and only query skipping is enabled (without synthesis). The total number of queries $N$ and the number of randomly generated queries $N_{\mathrm{init}}$ are $50$ and $1$. 

%%%%%%%%%%%%%%%%%%%%%%%%%%%%%%%%%%%%%%%%%%%%%%%%%%%%%%%%%%%%%%%%%%%%%%%%%%%%%%%%
\subsubsection{Simulated Users}
We designed several human preference models and prepared simulated users that answer queries according to these models. The simulated users have corresponding ${\bm w}_{(n)}^*$ and answer whether $\hat{\bf s}^{0}$ or $\hat{\bf s}^{1}$ is preferable by directly calculating the degree of preference $c_{(n)}$ from $\hat{\bf s}$ rather than ${\bf x}$, where the subscript $(n)$ indicates the $n$th user. Hence, they give low $c_{(n)}$ to the arrangements that do not satisfy the rules. To investigate how the uncertainty of answering affects the estimation, we prepared the following two types of simulated users:
\begin{itemize}
    \item \textbf{Ideal users}: return accurate answers based on \equref{equ: ideal_response}.
    \item \textbf{Uncertain users}: return uncertain answers based on \equref{equ: stupid_user_answer} in Appendix A. We prepared more uncertain users and less uncertain users that differ in the degree of uncertainty of their answers.
\end{itemize}
We also assume multiple users and prepared ${\bm w}^*_{(n)}$ for each type of simulated user: $0.1$, $0.5$, and $0.9$ for the simmered taro task and $(0.2, 0.6)$, $(0.3, 0.3)$, and $(1.0, 0.4)$ for the deep-fried shrimp task.
% We also prepared three ${\bm w}^*_{(n)}$ for each type of simulated user, $(0.2, 0.6)$, $(0.3, 0.3)$, and $(1.0, 0.4)$, assuming multiple users.

We ran ten trials under each condition. The parameters used in this experiments are available on our project page.

%%%%%%%%%%%%========================================================%%%%%%%%%%%%
\begin{figure}[t]
  \centering
  \begin{minipage}{0.49\linewidth}
    \centering
    \includegraphics[clip, trim=0 0 0 0, width=\linewidth]{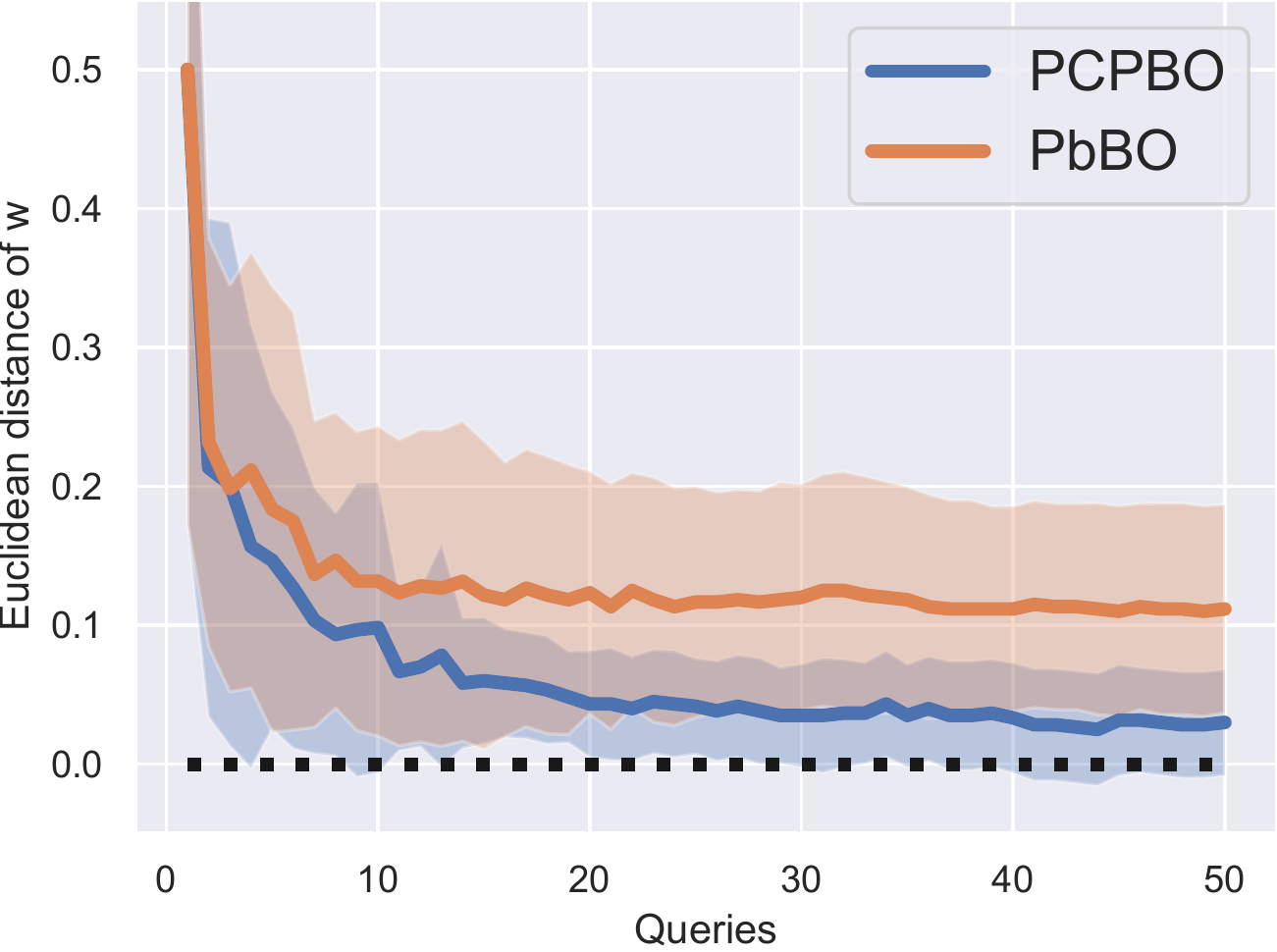}
    \subcaption{Simmered taro}
  \end{minipage}
  \begin{minipage}{0.49\linewidth}
    \centering
    \includegraphics[clip, trim=0 0 0 0, width=\linewidth]{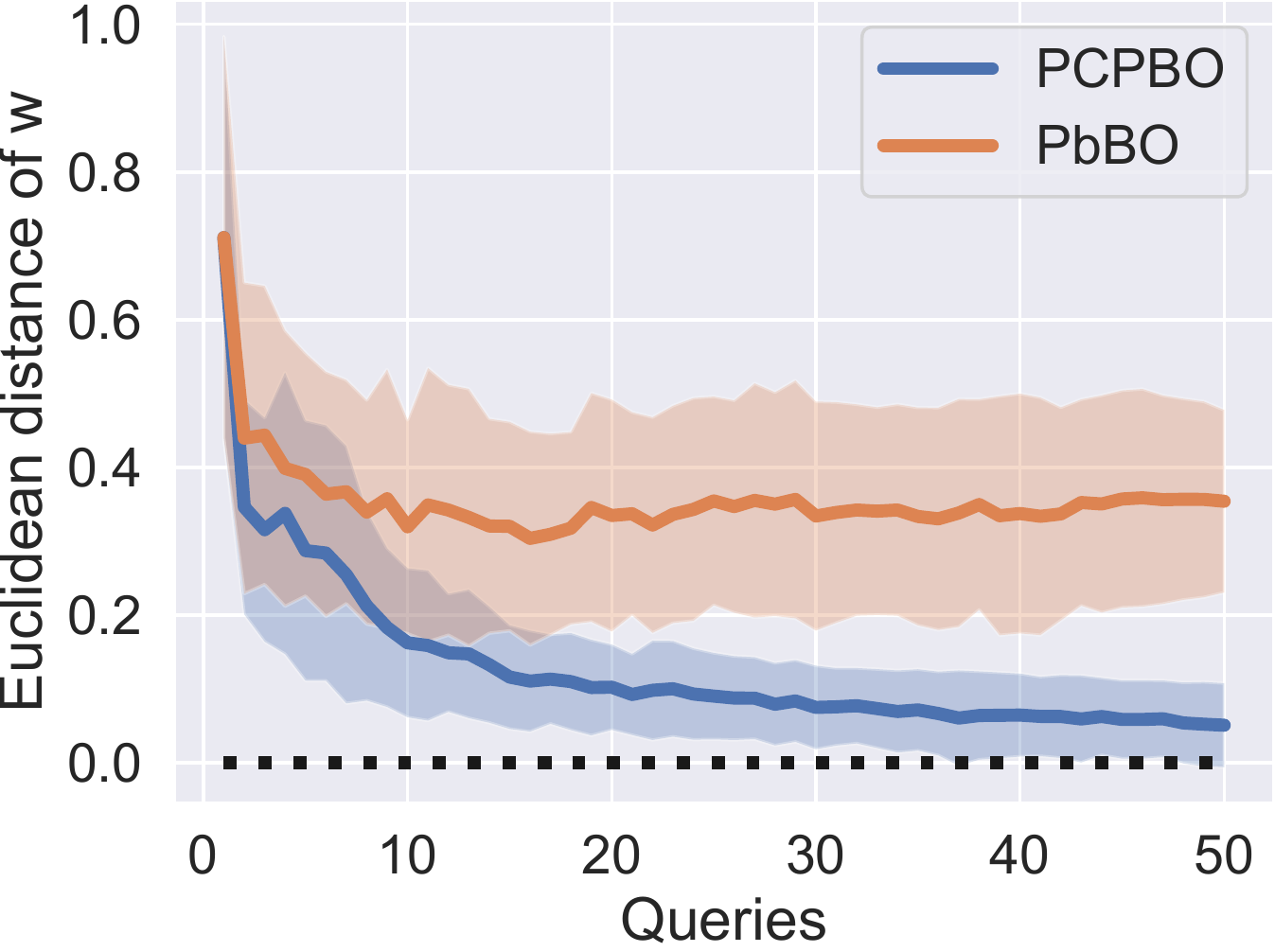}
    \subcaption{Deep-fried shrimp}
  \end{minipage}
  \caption{Transition of Euclidean distance between preferred weights and estimated weights according to number of queries for case of ideal users: Each solid line and shaded region are mean and standard deviation of ten trials of three simulated users.}
  \label{fig: sim_oracle_result}
\end{figure}
%%%%%%%%%%%%========================================================%%%%%%%%%%%%
%%%%%%%%%%%%========================================================%%%%%%%%%%%%
\begin{figure}[t]
  \centering
  \begin{minipage}{0.49\linewidth}
    \centering
    \includegraphics[clip, trim=0 0 0 0, width=\linewidth]{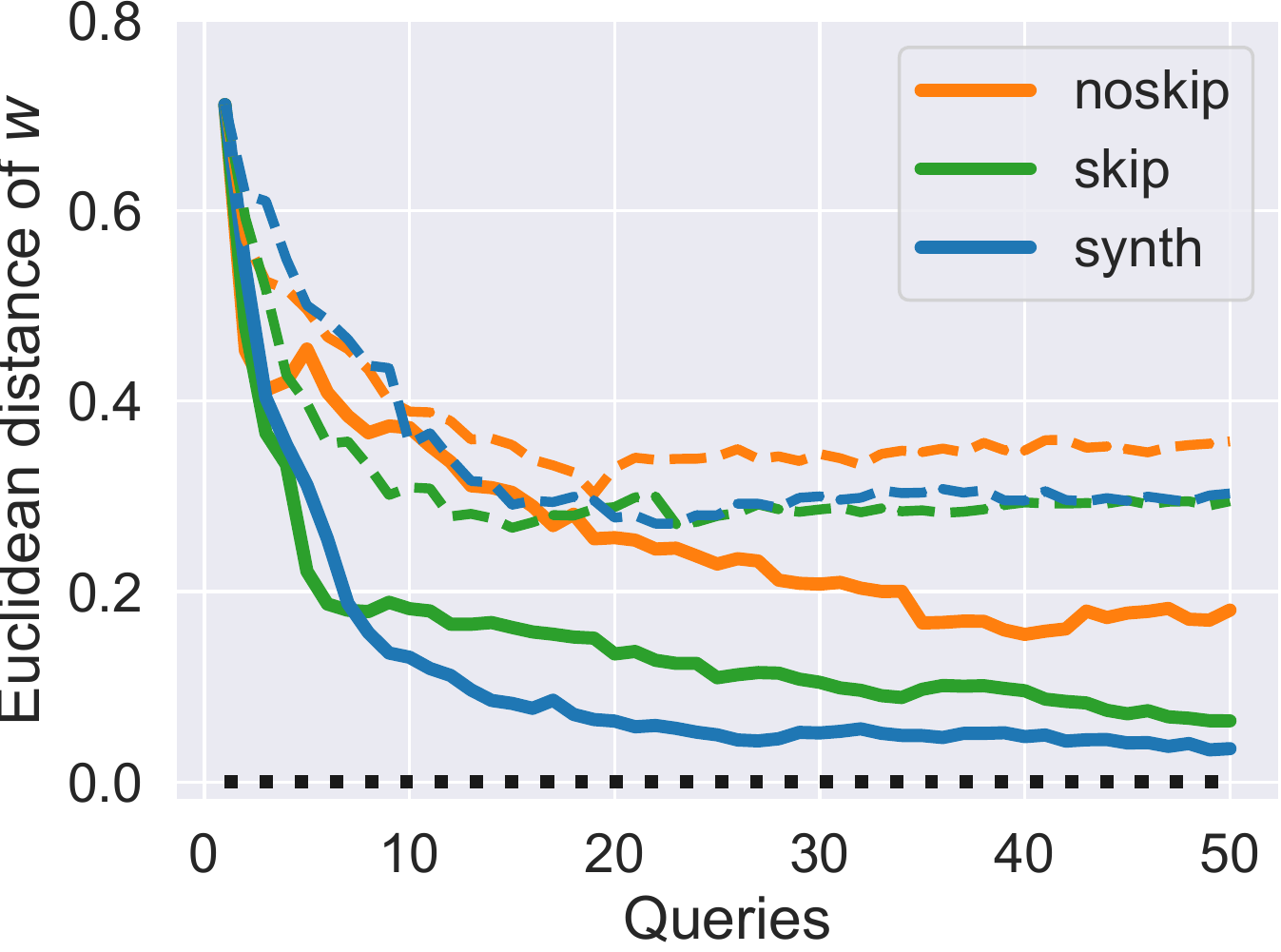}
    \subcaption{Less uncertain users}
  \end{minipage}
  \begin{minipage}{0.49\linewidth}
    \centering
    \includegraphics[clip, trim=0 0 0 0, width=\linewidth]{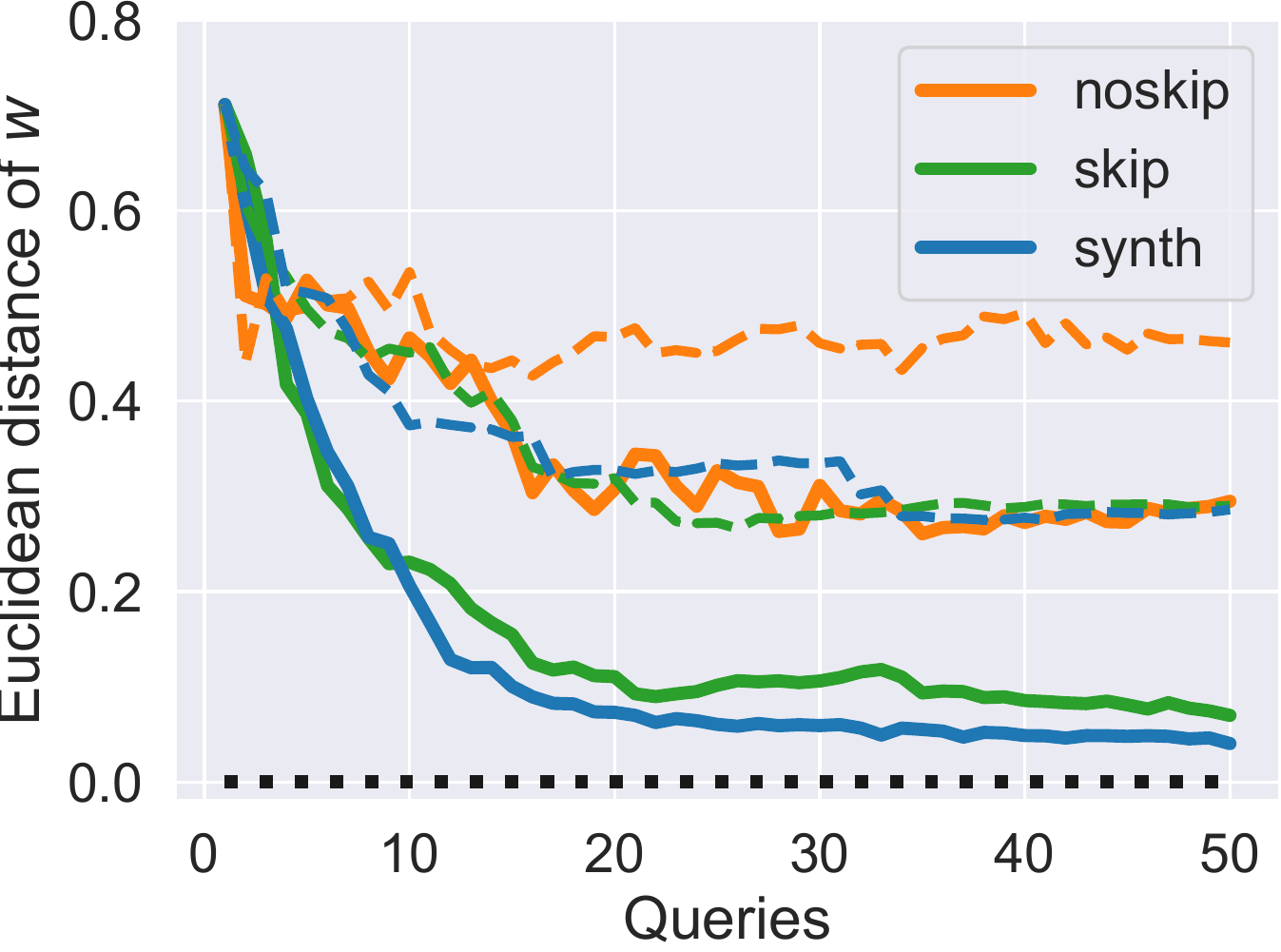}
    \subcaption{More uncertain users}
  \end{minipage}
  \caption{
  Comparison of preferred weight estimation results with and without query synthesis under uncertain users setting. Here $\texttt{noskip}$ indicates a setting w/o query synthesis, $\texttt{skip}$ denotes a setting w/ query skip (w/o synthesis), and $\texttt{syth}$ indicates a setting w/ query synthesis. Additionally, solid and dashed lines represent estimation results by PCPBO and PbBO.}
  \label{fig: sim_skip_result}
\end{figure}
%%%%%%%%%%%%========================================================%%%%%%%%%%%%

%%%%%%%%%%%%========================================================%%%%%%%%%%%%
\begin{figure}[t]
    \centering
    \includegraphics[width=\linewidth, clip]{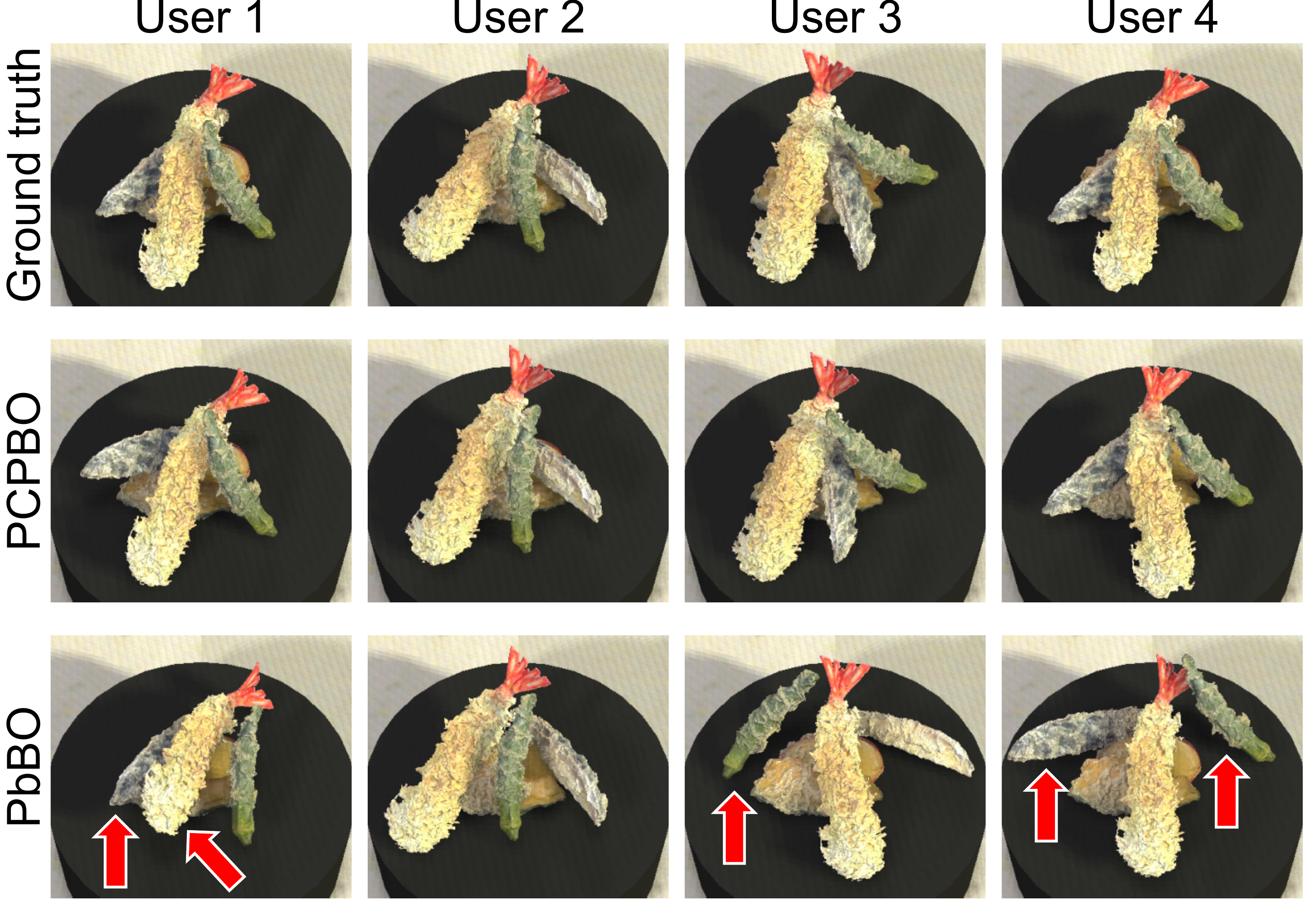}
    \caption{Comparison of pre-selected ($\texttt{Ground truth}$) and estimated arrangements ($\texttt{PCPBO}$ and $\texttt{PbBO}$) for each user: These figures are enlarged and clipped near center of original images. Red arrows indicate foods that fell over unexpectedly.}
    \label{fig: personal_estimate_img}
\end{figure}
%%%%%%%%%%%%========================================================%%%%%%%%%%%%
%%%%%%%%%%%%========================================================%%%%%%%%%%%%
\begin{figure}[t]
    \begin{minipage}[c]{0.485\linewidth}
        \includegraphics[width=\linewidth, clip, trim=0 0 0 0]{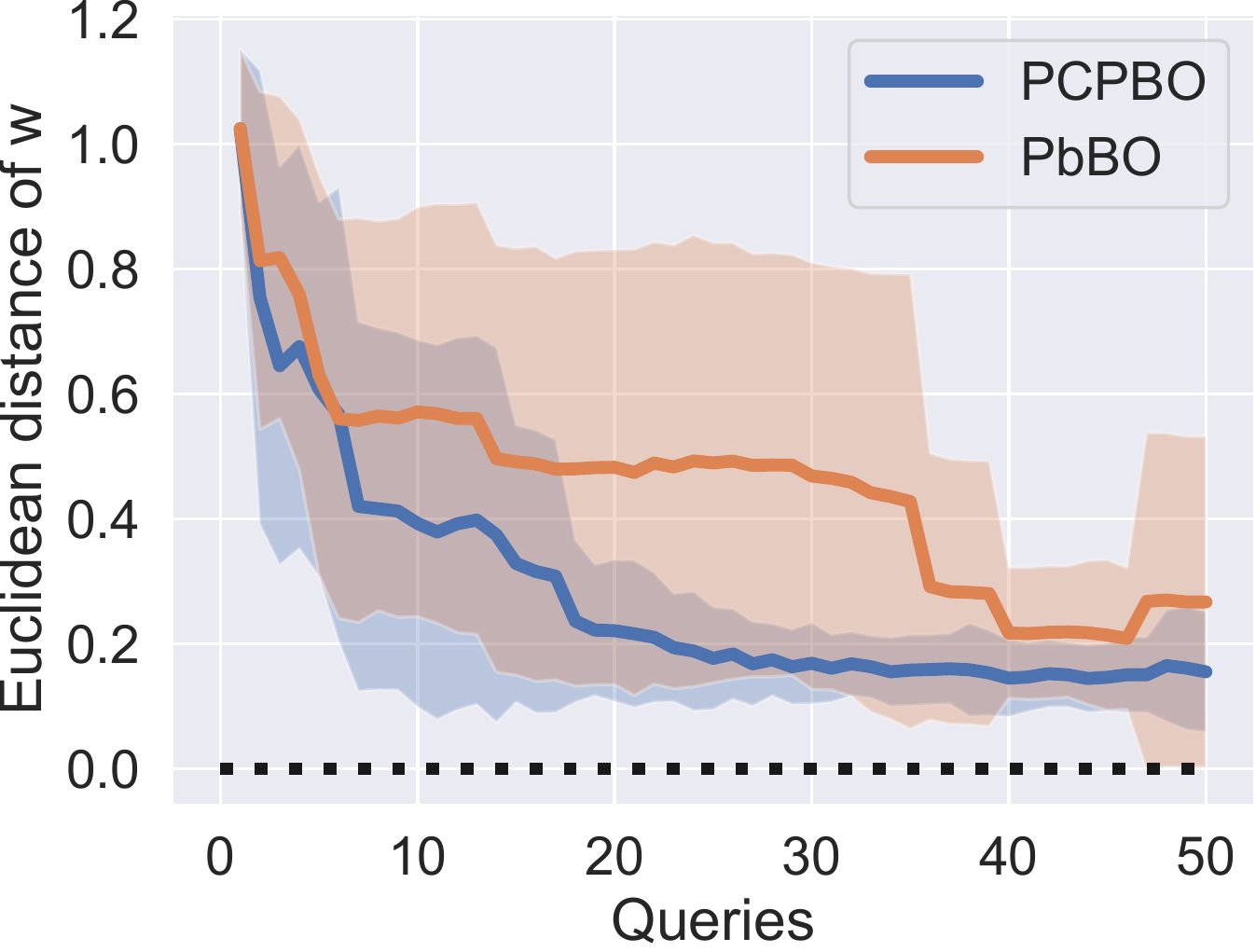}
    \end{minipage}\hfill
    \begin{minipage}[c]{0.46\linewidth}
        \caption{Transition of Euclidean distance between preferred weights and estimated weights according to number of queries for human users: Each solid line and shaded region are mean and standard deviation of three trials of four users.}
        \label{fig: estimate_w}
    \end{minipage}
\end{figure}
%%%%%%%%%%%%========================================================%%%%%%%%%%%%

%%%%%%%%%%%%%%%%%%%%%%%%%%%%%%%%%%%%%%%%%%%%%%%%%%%%%%%%%%%%%%%%%%%%%%%%%%%%%%%%
\subsection{Experimental Results}
%%%%%%%%%%%%%%%%%%%%%%%%%%%%%%%%%%%%%%%%%%%%%%%%%%%%%%%%%%%%%%%%%%%%%%%%%%%%%%%%
\subsubsection{Preferred Weight Estimation}
Figure \ref{fig: sim_oracle_result} shows the estimation results of ${\bm w}$ by PCPBO and PbBO under the ideal users setting. The horizontal axis represents the number of queries, and the vertical axis represents the Euclidean distance between ${\bm w}^*_{(n)}$ and estimated $\bm w$. Each solid line and the shaded region are the mean and standard deviation of ten trials for all the users. PCPBO, which considers physical consistency, estimates the preferred weights more accurately than physically inconsistent PbBO.

%%%%%%%%%%%%%%%%%%%%%%%%%%%%%%%%%%%%%%%%%%%%%%%%%%%%%%%%%%%%%%%%%%%%%%%%%%%%%%%%
\subsubsection{Analysis of Query Synthesis}
Figure \ref{fig: sim_skip_result} shows the estimation results of ${\bm w}$ under the uncertain users setting. $\texttt{synth}$, $\texttt{noskip}$, and $\texttt{skip}$ respectively represent the cases where query synthesis is enabled, disabled, and only query skipping is enabled (i.e., skipped queries are discarded). In $\texttt{synth}$ and $\texttt{skip}$, the users skipped the query in the third and fourth conditions on the right side of \equref{equ: stupid_user_answer} in Appendix A. Each solid and dashed line represents PCPBO and PbBO estimates, corresponding to the mean of ten trials for all the users. Forcing uncertain users to answer without providing the query skip mechanism lowers accuracy, which becomes more pronounced when uncertainty increases. The results of $\texttt{synth}$ are slightly better than that of $\texttt{skip}$. Finally, the skip rate's mean and standard deviation were $0.45 \pm 0.17$ for $\texttt{skip}$ and $0.25 \pm 0.12$ for $\texttt{synth}$ when answered by less uncertain users and $0.19 \pm 0.06$ for $\texttt{skip}$ and $0.12 \pm 0.04$ for $\texttt{synth}$ when answered by more uncertain users.

In summary, our proposed PCPBO estimates preferred weights with higher accuracy than the physically inconsistent PbBO when simulated users return accurate answers. Even when they return uncertain answers, PCPBO estimates precisely by combining the query synthesis.
%%%%%%%%%%%%%%%%%%%%%%%%%%%%%%%%%%%%%%%%%%%%%%%%%%%%%%%%%%%%%%%%%%%%%%%%%%%%%%%%
\section{User Study}
%%%%%%%%%%%%%%%%%%%%%%%%%%%%%%%%%%%%%%%%%%%%%%%%%%%%%%%%%%%%%%%%%%%%%%%%%%%%%%%%
\subsection{Experimental Setup}\label{subsec: hexp_setup}
Following our simulation experiment, we conducted an evaluation experiment with actual human users. This user study has three purposes. First, we compared the estimation performance of preferred weights between PCPBO and physically inconsistent PbBO. Second, we also compared the subjective preference of the results between them. Finally, we analyzed which queries were skipped by the users.

The experimental procedure is as follows: % divided into the following three parts:
\begin{enumerate}[\hspace{4mm}a)]
  \item confirm user's preferred arrangement (\ref{subsec: optimal_select})
  \item answer $N$ times two-choice queries (\ref{subsec: twochoice})
  \item answer likeability by a 7-point Likert scale (\ref{subsec: likert})
\end{enumerate}
where b) and c) were repeated six times for statistical evaluation. Here the first two times were performed in the order of PbBO and PCPBO, and the order of the remaining four was randomly determined.

%%%%%%%%%%%%%%%%%%%%%%%%%%%%%%%%%%%%%%%%%%%%%%%%%%%%%%%%%%%%%%%%%%%%%%%%%%%%%%%%
\subsubsection{Tempura Arrangement Task}
For this user study, we designed the arrangement task of tempura, which is a dish where ingredients are deep-fried. We chose sweet potato, pumpkin, eggplant, shrimp, and Japanese shishito pepper as ingredients. The task was designed from rules based on a Japanese chef's advice (\tabref{tab: knowledge}). Based on the rules, first we fixed the food's arrangement order and then set the sweet potato and pumpkin with specific states. We also restricted the states of the remaining foods and defined the states and actions as \equref{equ: exp_state} and \equref{equ: exp_action} and $\tau=({\bm s}_2, {\bm s}_3, {\bm s}_4)$ and $U=({\bm a}_2, {\bm a}_3, {\bm a}_4)$. The meshes for contact detection are approximated by rectangles.

%%%%%%%%%%%%========================================================%%%%%%%%%%%%
\begin{table}[t]
\caption{Subjective likeability of estimated arrangements by a 7-point Likert scale: Each method uses 50 times queries to perform estimation.}
\begin{tabular}{cc||cccc}
\hline
 & & \textbf{Trial 1} & \textbf{Trial 2} & \textbf{Trial 3} & {\textbf{Average score}} \\ \hline
\hline
\multirow{2}{*}{User 1} & PCPBO & 6 & 6 & 4 & $\bm{5.33 \pm 0.94}$          \\
                         & PbBO  & 2 & 5 & 4 & ${3.67 \pm 1.25}$              \\ \hline
\multirow{2}{*}{User 2} & PCPBO & 5 & 5 & 6 & ${5.33 \pm 0.47}$              \\ 
                         & PbBO  & 5 & 6 & 5 & ${5.33 \pm 0.47}$              \\ \hline
\multirow{2}{*}{User 3} & PCPBO & 5 & 6 & 5 & $\bm{5.33 \pm 0.47}$          \\ 
                         & PbBO  & 2 & 4 & 2 & ${2.67 \pm 0.94}$              \\ \hline
\multirow{2}{*}{User 4} & PCPBO & 6 & 7 & 3 & $\bm{5.33 \pm 1.70}$          \\ 
                         & PbBO  & 2 & 4 & 5 & ${3.67 \pm 1.25}$              \\ \hline
\end{tabular}
\label{tab: sel_likert}
\end{table}
%%%%%%%%%%%%========================================================%%%%%%%%%%%%
%%%%%%%%%%%%========================================================%%%%%%%%%%%%
\begin{table}[t]
\caption{Query skips per trial and average skip rate for each user}
    \centering
    \begin{tabular}{cc||cccc}
\hline
 & & \textbf{Trial 1} & \textbf{Trial 2} & \textbf{Trial 3} & {\textbf{Average rate}} \\ \hline 
\hline
\multirow{2}{*}{User 1}  & PCPBO & 11\, / \,50  & 23\, / \,50  & 11\, / \,50  & $\bm{0.30 \pm 0.11}$ \\ 
           & PbBO  & 38\, / \,50  & 21\, / \,50  & 43\, / \,50  & ${0.68 \pm 0.19}$ \\ \hline
\multirow{2}{*}{User 2}  & PCPBO & \ 5\, / \,50   & 14\, / \,50  & \ 7\, / \,50   & ${0.17 \pm 0.08}$ \\
           & PbBO  & \ 8\, / \,50   & \ 5\, / \,50   & \ 5\, / \,50   & $\bm{0.12 \pm 0.03}$ \\ \hline
\multirow{2}{*}{User 3}  & PCPBO & \ 8\, / \,50   & 11\, / \,50  & \ 4\, / \,50   & $\bm{0.15 \pm 0.06}$ \\
           & PbBO  & 21\, / \,50  & 23\, / \,50  & 15\, / \,50  & ${0.39 \pm 0.07}$ \\ \hline
\multirow{2}{*}{User 4}  & PCPBO & \ 0\, / \,50   & \ 2\, / \,50   & \ 0\, / \,50   & $\bm{0.01 \pm 0.02}$ \\
           & PbBO  & \ 8\, / \,50   & 10\, / \,50  & \ 3\, / \,50   & ${0.14 \pm 0.06}$ \\ \hline
\end{tabular}
    \label{tab: skip_rate}
\end{table}
%%%%%%%%%%%%========================================================%%%%%%%%%%%%
%%%%%%%%%%%%========================================================%%%%%%%%%%%%
\begin{figure}[t]
    \centering
    \includegraphics[width=0.9\linewidth, clip, trim=0 0 0 0]{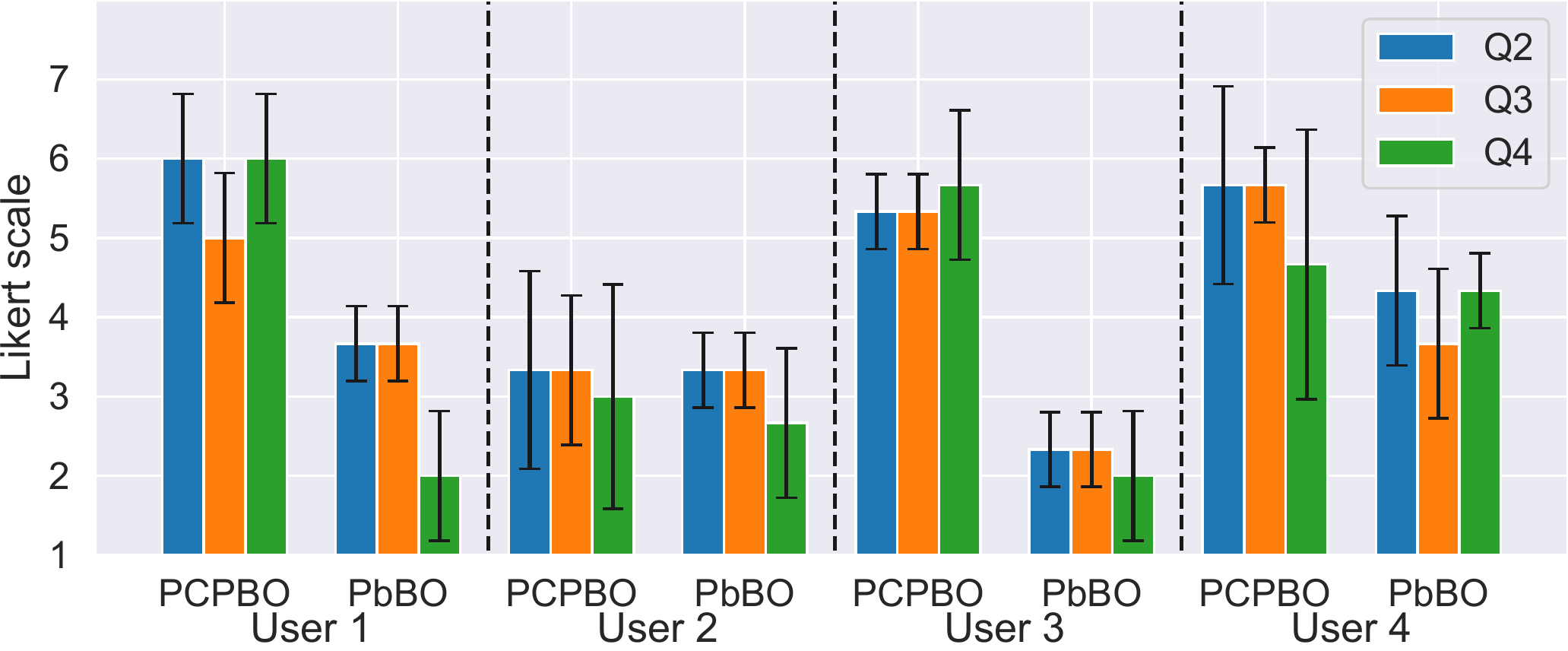}
    \caption{Results on subjective evaluation of quality of generated queries by a 7-point Likert scale: Each colored and error bar represents mean and standard deviation.}
    \label{fig: other_likert}
\end{figure}
%%%%%%%%%%%%========================================================%%%%%%%%%%%%
%%%%%%%%%%%%========================================================%%%%%%%%%%%%
\begin{figure}[t]
    \begin{minipage}[c]{0.49\linewidth}
        \includegraphics[width=\linewidth, clip, trim=0 0 0 0]{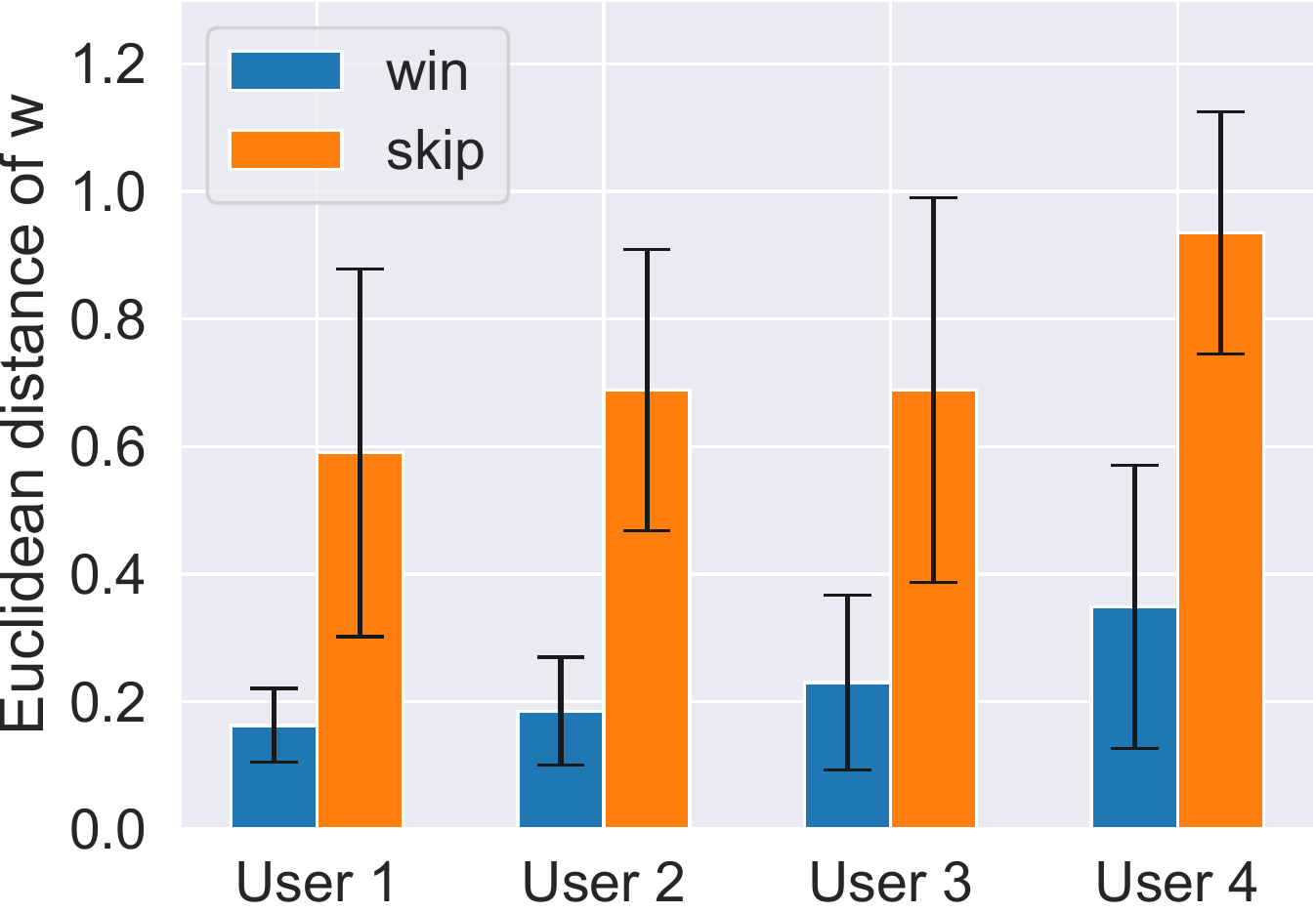}
    \end{minipage}\hfill
    \begin{minipage}[c]{0.46\linewidth}
        \caption{Comparison of Euclidean distances between skipped ($\texttt{skip}$) and unskipped ($\texttt{win}$) arrangements generated by PCPBO: $\texttt{skip}$ consists of the left and right side arrangements of skipped queries, and $\texttt{win}$ consists of the arrangement of chosen side in unskipped queries. Each colored and error bar represents mean and standard deviation.}
        \label{fig: skip_distance}
    \end{minipage}
    % \centering
\end{figure}
%%%%%%%%%%%%========================================================%%%%%%%%%%%%

%%%%%%%%%%%%%%%%%%%%%%%%%%%%%%%%%%%%%%%%%%%%%%%%%%%%%%%%%%%%%%%%%%%%%%%%%%%%%%%
\subsubsection{Preferred Arrangement Confirmation}\label{subsec: optimal_select}
The users selected the preferred arrangement in advance to quantitatively evaluate the distance between estimated $\bm w$ and ${\bm w}^*_{(n)}$. To shorten the experiment time, we ran CEM ten times for each $\bm w$ in advance to collect candidate images and presented the lowest cost image to the users.

%%%%%%%%%%%%%%%%%%%%%%%%%%%%%%%%%%%%%%%%%%%%%%%%%%%%%%%%%%%%%%%%%%%%%%%%%%%%%%%%
\subsubsection{Answering Two-Choice Queries}\label{subsec: twochoice}
The users answered $N$-time queries with ``I prefer the left one,'' ``I prefer the right one,'' and ``I don't like either.'' The pre-selected preferred arrangement was displayed near the user to allow quantitative evaluation. Although the arrangement must meet domain rules as the requirement of the foodservice industry, physically inconsistent PbBO often generates arrangements that violate them. Therefore, we taught the users the rules and instructed them to select ``I don't like either'' when neither arrangements were preferable or failed to satisfy the rules. To shorten the experiment time, CEM was not run in real-time in PCPBO, and the presenting images were randomly chosen from the candidate images collected in \ref{subsec: optimal_select}.

%%%%%%%%%%%%%%%%%%%%%%%%%%%%%%%%%%%%%%%%%%%%%%%%%%%%%%%%%%%%%%%%%%%%%%%%%%%%%%%%
\subsubsection{Evaluation by Likert Scale}\label{subsec: likert}
The users judged the likeability of the estimated arrangement images on the 7-point Likert scale where higher is more preferable (Q1). In addition, they answered the following three items:
\begin{itemize}
    \item Q2: quality of presented arrangements (the higher the better);
    \item Q3: ease of answering (the higher the easier);
    \item Q4: how quickly the preferred arrangement appeared (the higher the quicker).
\end{itemize}
At the end of the experiment, the users were allowed to revise their choices of the Likert scale for their consistency.

% As in the simulation experiment, PCPBO and PbBO were used to estimate the preferred weights.
PCPBO and PbBO searched for three-dimensional $\bm w$ space corresponding to $\psi_2$, $\psi_3$, and $\psi_4$, assuming the rules roughly determine $x$ and $y$ from $\psi$. Four men (age: 23--28) who are familiar with tempura participated in this user study. All the experiments were conducted under the approval of the Ethics Committee of Nara Institute of Science and Technology. The parameters used in this experiments are available on our project page.

%%%%%%%%%%%%%%%%%%%%%%%%%%%%%%%%%%%%%%%%%%%%%%%%%%%%%%%%%%%%%%%%%%%%%%%%%%%%%%%%
\subsection{Experimental Results}\label{subsec: hexp_result}
%%%%%%%%%%%%%%%%%%%%%%%%%%%%%%%%%%%%%%%%%%%%%%%%%%%%%%%%%%%%%%%%%%%%%%%%%%%%%%%%
\subsubsection{Preferred Weight Estimation}
Figure \ref{fig: personal_estimate_img} shows each user's pre-selected preferred arrangement (upper) and examples of the arrangement estimated by PCPBO (middle) and PbBO (lower). Here $\bm w^*_{(n)}$ is $(0.1, 0.4, 0.8)$, $(0.85, 0.15, 0.55)$, $(0.65, 0.4, 1.0)$, and $(0.1, 0.4, 0.85)$ in order from user 1. PCPBO's arrangements are less collapsed than PbBO's. Focusing on user 3, his pre-selected one and PbBO's are entirely different. On the other hand, there is little difference between the two estimated arrangements for user 2. This is because his pre-selected one is physically stable, and it is sufficient to use the map $d$, which only considers the rules. Therefore, we note that the results for user 2 tend to differ from those of the other users. Next \figref{fig: estimate_w} shows the estimation results of ${\bm w}$ by PCPBO and PbBO. The horizontal axis represents the number of queries, and the vertical axis represents the Euclidean distance between ${\bm w}^*_{(n)}$ and estimated $\bm w$. Each solid line and the shaded region are the mean and standard deviation of three trials for four users. These results indicate that PCPBO estimated ${\bm w}^*_{(n)}$ more accurately and stably than PbBO.

%%%%%%%%%%%%%%%%%%%%%%%%%%%%%%%%%%%%%%%%%%%%%%%%%%%%%%%%%%%%%%%%%%%%%%%%%%%%%%%%
\subsubsection{Subjective Preference of Queries}
Table \ref{tab: sel_likert} shows the results of asking about the likeability for the estimated arrangement images on the 7-point Likert scale. PCPBO's arrangement tended to be preferred over PbBO's, except by user 2, and the paired t-test results indicate a significant difference between them ($p=0.02$). Next \figref{fig: other_likert} shows the Likert scale results for the three items described in \ref{subsec: likert}. PCPBO's scores are equal to or higher than PbBO's in the subjective ratings, and the paired t-test results for each item are ($p=0.009$) for Q2, ($p=0.002$) for Q3, and ($p=0.013$) for Q4, indicating a significant difference between them. These results show that PCPBO's arrangement is also more preferred in subjective evaluations.

%%%%%%%%%%%%%%%%%%%%%%%%%%%%%%%%%%%%%%%%%%%%%%%%%%%%%%%%%%%%%%%%%%%%%%%%%%%%%%%%
\subsubsection{Analysis of Query Synthesis}
Table \ref{tab: skip_rate} compares PCPBO and PbBO in terms of query skipping. The average skip rate for PCPBO is lower except for user 2, which suggests that it generates higher quality queries than PbBO. Figure \ref{fig: skip_distance} compares the distances of ${\bm w}$ that corresponds to the skipped and unskipped arrangements in the queries generated by PCPBO. In \figref{fig: skip_distance}, $\texttt{win}$ corresponds to the set of selected arrangements by the user in the unskipped queries, and $\texttt{skip}$ corresponds to the set of the left and right arrangements in the skipped queries. The users properly skipped the unpreferable ones from the $\texttt{skip}$ results that are relatively far apart; the query synthesis properly worked well even for actual human users.

In summary, even with human users, our proposed PCPBO qualitatively and quantitatively outperformed the physically inconsistent PbBO. Moreover, these results show that the query synthesis framework worked as intended.

%%%%%%%%%%%%%%%%%%%%%%%%%%%%%%%%%%%%%%%%%%%%%%%%%%%%%%%%%%%%%%%%%%%%%%%%%%%%%%%%
\section{Discussion}
Although this paper focused on Japanese cuisine, PCPBO is likely applicable to various cuisines from other countries. However, our proposed method has some limitations. First, our experiments approximated the foods as rigid bodies with simple shapes. This simplification will be an obstacle when considering a robotic food arrangement. We might need a better solution, e.g., a better physical simulator and an approximation method of foods\cite{residual_model}. Next, a cost function and map corresponding to $\bm w$ must be implemented by interpreting the rules, which is tedious. It would be valuable to address this difficulty with machine learning approaches. As another future direction, we will consider integrating two-choice queries and demonstrations\cite{Matsuoka} to estimate the preferred arrangement with fewer interactions. 
In addition, since the proposed method estimates the preferred object arrangement while simultaneously considering domain rules and physical constraints, it may be applicable to palletizing tasks\cite{palletizing} in the last mile, where driver preferences need to be considered, and preference-based cleanup tasks\cite{tydingup}. To those end, like our tempura case, 3D scanning of the environment and objects in the task would be needed to prepare a physical simulator with CG. Furthermore, in the former task, it is difficult for the users to estimate the weight of cardboard boxes from visual information alone, so we may need to handle weight data explicitly. Also, the latter task may require extending our method for handling many objects, including unknown ones.
%%%%%%%%%%%%%%%%%%%%%%%%%%%%%%%%%%%%%%%%%%%%%%%%%%%%%%%%%%%%%%%%%%%%%%%%%%%%%%%%
\section{Conclusion}
We proposed Physically Consistent Preferential Bayesian Optimization (PCPBO) for estimating the preferred food arrangement that satisfies domain rules and physical laws. The results of simulator-based tasks with simulated and human users show that our method can estimate user-preferred arrangements in which both are taken into account.

%%%%%%%%%%%%%%%%%%%%%%%%%%%%%%%%%%%%%%%%%%%%%%%%%%%%%%%%%%%%%%%%%%%%%%%%%%%%%%%%
\section*{APPENDIX}
%%%%%%%%%%%%%%%%%%%%%%%%%%%%%%%%%%%%%%%%%%%%%%%%%%%%%%%%%%%%%%%%%%%%%%%%%%%%%%%%
\subsection{Uncertain User Model}
Let two thresholds be $t_0$ and $t_1$ $(t_0 < t_1)$, and if $c_{\mathrm{min}}=\mathrm{\mathrm{min}}(c^0_{(n)}, c^1_{(n)})$, uncertain users return an uncertain answer as follows:
\begin{equation}
    y = \begin{cases}
    H(c^1_{(n)}-c^0_{(n)}), & \text{if $c_{\mathrm{min}} < t_0$}\\
    H(c^1_{(n)}-c^0_{(n)}), & \text{if $t_0 \leq c_{\mathrm{min}} \leq t_1$}\\
     & \hspace{4mm} \text{and $\frac{c_{\mathrm{min}} - t_0}{t_1 - t_0} \leq \mathrm{rand}()$}\\
    H(\mathrm{rand}()-0.5), & \text{if $t_0 \leq c_{\mathrm{min}} \leq t_1$}\\
     & \hspace{4mm} \text{and $\frac{c_{\mathrm{min}} - t_0}{t_1 - t_0} > \mathrm{rand}()$}\\
    H(\mathrm{rand}()-0.5), & \text{if $c_{\mathrm{min}} > t_1$}
    \end{cases}
    \label{equ: stupid_user_answer}
\end{equation}
where $H(\cdot)$ is a unit step function and $\mathrm{rand}()$ returns a uniform random number within $[0,1]$. To change the uncertainty of answering, we prepared two $t_1$s, $50$ (less uncertain users) and $100$ (more uncertain users). Also, $t_0$ was set to $20$.

\bibliographystyle{IEEEtran}
\bibliography{kyhw_ref}
%%%%%%%%%%%%%%%%%%%%%%%%%%%%%%%%%%%%%%%%%%%%%%%%%%%%%%%%%%%%%%%%%%%%%%%%%%%%%%%%
%%%%%%%%%%%%%%%%%%%%%%%%%%%%%%%%%%%%%%%%%%%%%%%%%%%%%%%%%%%%%%%%%%%%%%%%%%%%%%%%
% \clearpage
% \section*{supplementary material}}
% \input{tex_files/99_sup}
\end{document}